# Magnetization and Polarization of Coupled Nuclear Spins Ensembles

Danila A. Barskiy[1] and Andrey N. Pravdivtsev[2]

1 - Institut für Physik, Johannes Gutenberg Universität Mainz, 55128 Mainz, Germany; Helmholtz Institut Mainz, 55128 Mainz, Germany; GSI Helmholtzzentrum für Schwerionenforschung, Darmstadt, Germany. E-mail: dbarskiy@uni-mainz.de

2 - Section Biomedical Imaging, Molecular Imaging North Competence Center (MOIN CC), Department of Radiology and Neuroradiology, University Medical Center Kiel, Kiel University, Am Botanischen Garten 14, 24118, Kiel, Germany. E-mail: andrey.pravdivtsev@rad.uni-kiel.de

**Abstract**

In nuclear magnetic resonance (NMR), the bulk magnetization of a sample is commonly assumed to be proportional to spin polarization, with each spin of the same type contributing equally to the measured signal. In this work, we prove the *high-field theorem* for general spin-*I* systems (where *I* is the spin quantum number): the total measurable NMR signal remains unaffected by the grouping of spins into equivalent units (e.g., molecules), provided the system is at thermodynamic equilibrium in the high field limit ($\hbar\omega_0 \gg |H_{\mathrm{spin-spin}}|$, where $\omega_0$ is the Larmor frequency and $|H_{\mathrm{spin-spin}}|$ characterizes internal spin-spin interactions). The results are derived using both magnetization equations and density matrix formalism. The theorem, however, does not extend to conditions far from thermodynamic equilibrium or such as zero- to ultralow-field NMR. We also present three educational problems designed to deepen understanding of the material in classroom settings. This work reinforces established principles in magnetic resonance but also highlights areas for further exploration.

## Introduction

Polarization is an important property of spin ensembles. Polarization corresponds directly to the measured signal in experimental techniques like nuclear magnetic resonance (NMR) [1], its imaging version, magnetic resonance imaging (MRI), and electron paramagnetic resonance (EPR) [2,3]. At a typical magnetic field of 1 T and room temperature, the polarization of $^{1}$H nuclei is on the order of $10^{-5}$ (0.001%). Because of the direct proportionality of magnetic resonance signal to polarization, approaches to increase polarization (to create the so-called "hyperpolarization") are actively being developed [4]. In various textbooks, scientific reviews, and peer-reviewed publications [5], polarization is typically defined for an ensemble of uncoupled spin-½ particles[1]. For such a two-level system, polarization along the direction of the magnetic field ($P_{\mathrm{z}}$, superscript {1/2} indicates spin quantum number) is readily defined by the following equation:

$$P_{\mathrm{z}}^{\{1/2\}} = \frac{n_\alpha - n_\beta}{n_\alpha + n_\beta}. \quad (1)$$

Here, $n_\alpha$ and $n_\beta$ are populations of the spin states that correspond to spin orientations along the magnetic field ($\alpha$) and in the opposite direction ($\beta$), respectively. This paper will focus on analyzing *nuclear* spin ensembles and corresponding polarization values.

It is important to note that despite nuclear spins can be both fermions (half-odd-integer spins, for example, $^{1}$H, $^{13}$C, $^{15}$N, $^{19}$F, $^{31}$P) and bosons (integer spins, for example, $^{2}$D, $^{6}$Li, $^{14}$N, $^{36}$Cl), in virtually all situations encountered in solution-state NMR at standard conditions (~25 $^{o}$C, ~1 bar) nuclear wavefunctions do not overlap, and Boltzmann statistics can be applied for both types of particles [6,7]. Therefore, by defining the Boltzmann factor $\mathbb{B}$ as a ratio of nuclear energy-level splitting due to interaction of spins with the external $B_{\mathrm{z}}$ is magnetic field (**Figure 1**) and an available thermal energy in the system, $\mathbb{B} = \gamma\hbar B_{\mathrm{z}}/k_{\mathrm{B}}T$ (where $\hbar$ and $k_{\mathrm{B}}$ are Plank's and Boltzmann's constants, respectively), polarization of the ensemble of non-interacting spins-½ ($P_{\mathrm{z}}^{\{1/2\}}$) at thermodynamic equilibrium follows from the eq. (1):

$$P_{\mathrm{z}}^{\{1/2\}} = \frac{e^{\mathbb{B}/2} - e^{-\mathbb{B}/2}}{e^{\mathbb{B}/2} + e^{-\mathbb{B}/2}} = \tanh(\mathbb{B}/2). \quad (2)$$

At thermodynamic equilibrium under high temperature (HT) approximation (i.e., when $\mathbb{B} \ll 1$), $e^{\pm\mathbb{B}/2} \approx 1 \pm \mathbb{B}/2$, thus,

$$P_{\mathrm{z}}^{\{1/2\}} \cong (1/2) \cdot \mathbb{B}. \quad (3)$$

One can readily see that for a specific case of isolated spins-½, overpopulation of one level with respect to another creates *polarization*, i.e., a specific *orientation* of a vector quantity in space. Therefore, restricting "polarization" and "hyperpolarization" to describe vector quantities seems reasonable. However, Bengs and Levitt recently presented compelling arguments as to why this terminology can also be used to represent other spin orders [8]. At the same time, a specific distribution of populations could also have different names, such as alignment, scalar order, population imbalance, latent polarization, quadrupolar order etc. [8–11].

[1]By particle we mean either individual nuclei or a collection of them. For example, an ensemble of nuclear spins in a molecule with total spin-*J* can be referred to as spin-*J* quasi-particle.

Our manuscript is motivated by trying to answer two seemingly simple questions: (i) What is polarization for a general case of spin-*I* particles at thermal equilibrium? (ii) Does the magnetic equivalence of spins in molecules nonlinearly affect total magnetization of the system at thermal equilibrium?

In the following, we will consider these questions and demonstrate several approaches to finding the answers. We constrain ourselves to situations of nuclear spin systems at thermodynamic equilibrium and high magnetic fields, i.e., when Zeeman interactions are much larger than any other interactions in the system, $\gamma\hbar B_{\mathrm{z}} \gg \left\|H_{\mathrm{spin-spin}}\right\|$. In this case, one can neglect all interactions except Zeeman in spin state population distribution.

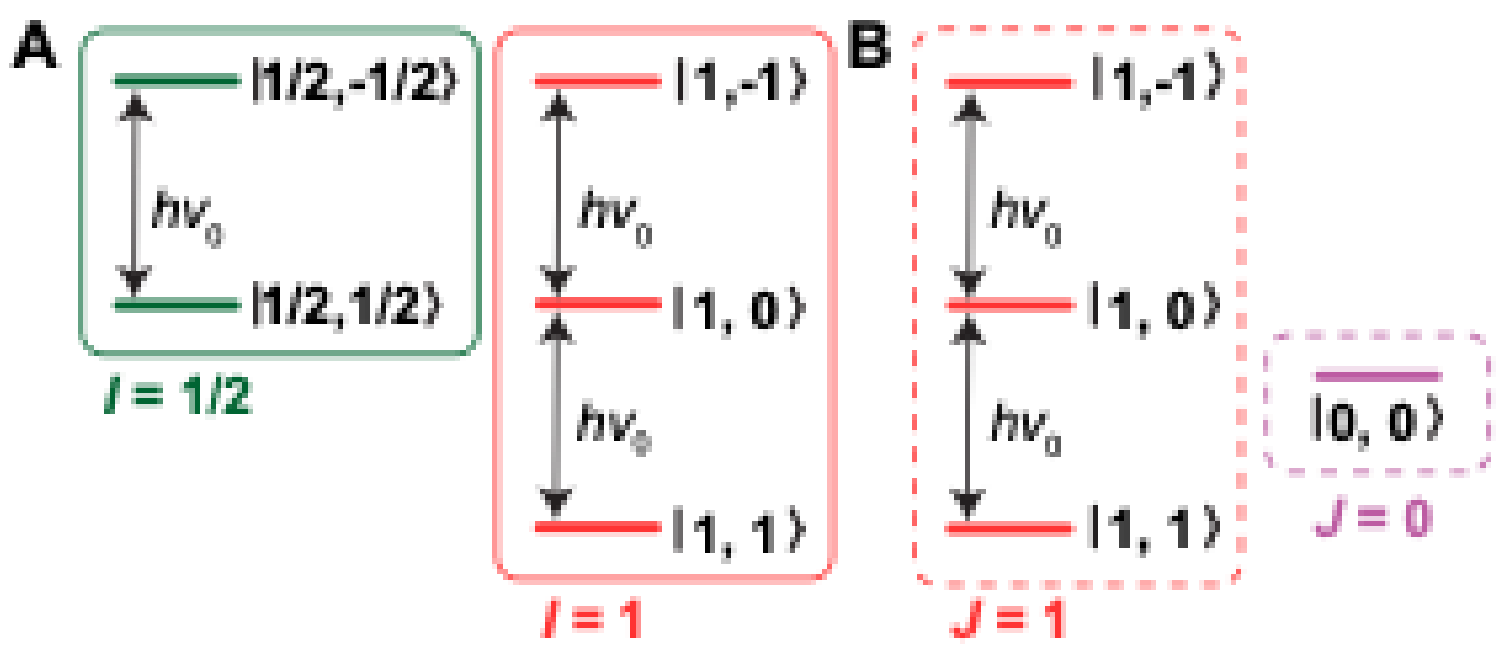

**Figure 1.** Energy-level diagrams for (A) spin-½ and spin-1 particles and (B) two equivalent spins-½ (allowed NMR transitions are shown with frequency $\nu_0 = |\gamma B_0/2\pi|$ ($\gamma > 0$ is assumed). Notice that in (B), two spins are treated as two semi-independent spin-1 and spin-0 quasi-particles.

# Results and discussion

### Formal definition of polarization from bulk magnetization

Magnetic field $\vec{B}$ in a material is typically defined as a sum of two fields, $\vec{H}$ and $\vec{M}$ such that $\vec{B} = \mu_0\left(\vec{H} + \vec{M}\right)$, where $\vec{M}$ is magnetization (the dimension of the components in MKS units is [A/m]). When referring to magnetization in this work, we consider it to be created by nuclear spins. Assuming homogeneous magnetization in volume *V*, the magnetic moment $m_{\mathrm{z}}$ of the whole sample (units [A·m$^2$]) is defined as $m_{\mathrm{z}}^{\{I\},N} = M_{\mathrm{z}} \cdot V$. The superscript "$\{I\}, N$" indicates that magnetization is generated by *N* spin-*I* particels:

$$m_{\mathrm{z}}^{\{I\},N} = N \cdot \langle \mu_{\mathrm{z}}^{I} \rangle = N \cdot \gamma\hbar \langle I_{\mathrm{z}} \rangle. \tag{4}$$

Here, $\langle \mu_{\mathrm{z}}^{I} \rangle$ is an average magnetic moment per spin-*I* measured along the *z*-axis. For a single spin, $\mu_{\mathrm{z}}^{I}$ is associated with spin projection $I_{\mathrm{z}}$ and its behavior is theoretically described by a wavefunction $|\psi\rangle = |I, I_{\mathrm{z}}\rangle$. Eq. (4) can be considered "Lemma 0" of this work; while it is trivial, it formally defines $\langle \mu_{\mathrm{z}}^{I} \rangle$.

Since for a fully (100%) polarized system of spin-*I* particles the maximum magnetic moment is reached when all spins have their maximal projection $\langle \mu_{\mathrm{z}}^{I} \rangle_{\mathrm{max}} = \gamma\hbar I$,

$$m_{\mathrm{z,max}}^{\{I\},N} = N \cdot \gamma\hbar I. \tag{5}$$

For a fully polarized spin-*I* system, the state with projection *I* (or –*I*) has a population of 1. Note that spin-0 particles have no sizable projection of spin in any direction, hence, they do not induce magnetization.

In all other cases, it is reasonable to define polarization, $P_{\mathrm{z}}^{\{I\}}$, as a *z*-component of a vector quantity proportional to magnetization normalized to its maximal value [12]:

$$P_{\mathrm{z}}^{\{I\}} = \frac{m_{\mathrm{z}}^{\{I\},N}}{m_{\mathrm{z,max}}^{\{I\},N}} = \frac{\langle I_{\mathrm{z}} \rangle}{I} = \frac{1}{I} \sum_{I_{\mathrm{z}}=I}^{-I} I_{\mathrm{z}} \cdot \rho_{I_{\mathrm{z}}}. \tag{6}$$

Here, $\rho_{I_{\mathrm{z}}}$ is a fraction of spins with projection $I_{\mathrm{z}}$; summation runs over $(2I+1)$ states, from $|I, I\rangle$ to $|I, -I\rangle$. Note that if magnetization was measured in directions other than *z*, the basis would have to change accordingly. By using a density matrix operator, $\hat{\rho}$, instead of wave functions, the same information about the system can be obtained [10]:

$$P_{\mathrm{z}}^{\{I\}} = \frac{1}{I} \mathrm{Tr}\big(\hat{I}_{\mathrm{z}} \cdot \hat{\rho}\big). \tag{7}$$

We assume that the density matrix is normalized, i.e., $\mathrm{Tr}(\hat{\rho}) = 1$. In cases where the density matrix is diagonal in the basis corresponding to the measurement basis, eqs. (6) and (7) are identical—otherwise, eq. (7) is more general than eq. (6) as it is valid in any basis set. Since we excluded cases of zero spin, one may notice that $P_{\mathrm{z}}^{\{I\}}$ can be always factorized out of the average magnetic moment per particle, therefore, it represents an *ensemble property* and does not depend on *N*.

To summarize and generalize eq. (4) to the ensemble of different spins, we conclude that the magnetic moment of a sample can always be represented as a product of the number of particles times maximal magnetic moment per particle times polarization of the ensemble:

$$m_{\mathrm{z}}^{\{I_k\},N_k} = N_k \cdot \langle \mu_{\mathrm{z}}^{I_k} \rangle \; = N_k \cdot (\gamma_k \hbar I_k) \cdot P_{\mathrm{z}}^{\{I_k\}}. \tag{8}$$

Here, $I_k$ is a quantum number corresponding to the *k*-th spin type (1/2, 3/2 etc.) with gyromagnetic ratio $\gamma_k$ and $N_k$ is the number of *k*-th spins in the sample. A “particle” can represent a spin, a molecule, or even a specific realization of total spin made by a group of nuclei in the molecule (quasi-particles).

## Thermodynamic Equilibrium, High-Field and High-Temperature Approximations

In thermodynamic equilibrium, a partition function $\mathbb{Z}$ and temperature *T* are defined via average energy $\langle E \rangle$ of the system such that

$$\langle E \rangle = -kT \ln \mathbb{Z}. \tag{9}$$

Since for spins $\langle E \rangle = -(\gamma \hbar I) P_{\mathrm{z}}^{\{I\}} B_{\mathrm{z}}$, polarization can be deduced from the partition function:

$$P_{\mathrm{z}}^{\{I\}} = \frac{1}{I} \frac{d \ln \mathbb{Z}}{d \mathbb{B}}. \tag{10}$$

If a spin system is in a state that is far from thermodynamic equilibrium eqs. (9-10) cannot be used.

In this work, we will only consider high-field (HF) approximation: the case when the interaction of a spin-*I* system with the external magnetic field dominates all other spin-spin interactions, $\gamma \hbar B_{\mathrm{z}} \gg \left\| H_{\mathrm{spin-spin}} \right\|$. Under this assumption, the partition function has a simple exponential structure:

$$\mathbb{Z} \approx \sum_{I_z} e^{\mathbb{B} I_z}. \tag{11}$$

In the case of additional "high-temperature" (HT) approximation, a further condition is set: the thermal energy of the system is larger than gaps between spin energy levels: $kT \gg \gamma\hbar B_z \gg \|H_{\text{spin-spin}}\|$ and exponentials can be simplified using Taylor expansion:

$$\mathbb{Z}^{\text{HT}} = \sum_{I_z} (1 + \mathbb{B} I_z) + o(\mathbb{B}). \tag{12}$$

In the following, we only consider cases of thermodynamic equilibrium and HF approximation and omit additional sub/superscripts to indicate this. When HT approximation is used, it will be marked with a superscript 'HT'.

### Polarization of a General Spin-*I* System

Following eqs. (6-7) and/or eqs. (10-11), polarization at thermal equilibrium and at high fields can be calculated as

$$P_z^{\{I\}} = \frac{\sum_{I_z} I_z e^{\mathbb{B} I_z}}{I \cdot \sum_{I_z} e^{\mathbb{B} I_z}}. \tag{13}$$

To obtain an analytical expression, one can use known sums of geometric progression (see **Appendix A1**):

$$P_z^{\{I\}} = \frac{1}{I}\left\{\left(I + \frac{1}{2}\right)\coth\left(\left(I + \frac{1}{2}\right)\mathbb{B}\right) - \frac{1}{2}\coth\left(\frac{\mathbb{B}}{2}\right)\right\}, \tag{14}$$

which is known as the Brillouin function, $P_z^{\{I\}} \in [-1,1]$ [12]. Under HT approximation ($\mathbb{B} \to 0$) polarization simplifies to

$$P_z^{\{I\},\text{HT}} \cong \frac{I + 1}{3}\mathbb{B}. \tag{15}$$

This result can be also obtained directly from eq. (6) considering the population of a state with a projection $I_z$ is $(1 + \mathbb{B} I_z)/(2I + 1)$. Then, using the known equalities for a sum of squares, including the case of half-integer numbers (**Appendix A2**), eq. (15) is derived.

With this, we answer question (i) laid out in the introduction by finding polarization of a system of independent spins-*I* given HF approximation.

## Magnetization of an arbitrary spin system

Having found the polarization of spin-I at thermal equilibrium, we will now see if it is different when it is one of many coupled equivalent spins. This we will do in 2 steps, formulating two lemmas and then a general theorem for the magnetization of spins.

**Lemma 1**: Under high-field conditions, the magnetization of a system consisting of *k* different types of magnetically inequivalent spins is a sum of their individual magnetizations:

$$m_z^{\{I_1 I_2 \ldots I_k\},N} = \sum_{r=1}^{k} m_z^{\{I_r\},N_r}. \tag{16}$$

Here, $N_k$ is a number of spins of the type $k$, and the sum runs over all possible *k* types of spins, $N = \sum_{r=1}^{k} N_r$. The syntaxis above implies that $m_z^{\{I_r\},N_r}$ corresponds to the magnetic moment originating from only one type of spins, spin-$I_r$ particles. Proof comes from the fact that the magnetic field is an additive property of its sources; this is generally true when the field originating from each spin is small compared to the external magnetic field. Otherwise, one should consider additional field amplification.

**Lemma 2.** Under high-field thermal equilibrium conditions, the magnetization of $N^{(i)}$ molecules of the type (*i*)---each containing $n^{(i)}$ magnetically equivalent spins-*I* equals to $n^{(i)}N^{(i)}$ magnetizations of non-interacting spins-*I*:

$$m_z^{\{n^{(i)}\times I\},N^{(i)}} = n^{(i)} \cdot m_z^{\{I\},N^{(i)}} = n^{(i)} N^{(i)} \cdot \langle \mu_z^{I} \rangle. \tag{17}$$

The proof is given in **Appendixes A4-5** using an approach involving classical magnetization and **Appendix A6** using the density matrix treatment. Eq. (17) can also be seen as an extension to the indexing notation introduced above (eq. 8). Collorary of the Lemma 2 is that grouping of spins in molecules does not affect the total magnetization of the system, i.e., $m_z^{\{n\times I\},N/n} = m_z^{\{I\},N} = N \cdot (\gamma \hbar I) \cdot P_z^{\{I\}}$. This answers the question (ii) laid out in the introduction.

**Theorem 1.** Magnetization of any spin system at thermal equilibrium and at a high field ($\gamma\hbar B_z \gg \|H_{\text{spin-spin}}\|$) is an additive quantity that can be expressed as

$$m_z^{\{I_1 I_2 \ldots I_k\},N} = \sum_i m_z^{\{n_1^{(i)}\times I_1,\ldots,n_k^{(i)}\times I_k\},N^{(i)}} = \sum_{r=1}^{k} N_r \, \langle \mu_z^{I_r} \rangle. \tag{18}$$

Here, $n_k^{(i)}$ is a number of spins of the type *k* in molecules of the type $(i)$. Summation over all molecules gives a number $N_r = \sum_i \, n_r^{(i)} N^{(i)}$ of the *r*-th type of spins in the system and $\langle \mu_z^{I_r} \rangle = (\gamma \hbar I_r) \cdot P_z^{\{I_r\}}$.

The theorem follows directly from **Lemmas 1** and **2**. To summarize, one can see that at thermal equilibrium in high field, it is always possible to divide total magnetization into individual contributions originating from different spin types and/or different molecules, irrespective of their magnetic equivalence (**Figure 2**).

While these properties (eq. 16-18) may seem obvious, it is essential to realize that conditions of their applicability do not hold at low fields (i.e., when internal spin-spin interactions dominate Zeeman interactions) and far from thermodynamic equilibrium.

This theorem has profound implications in conventional high-field NMR experiments. Indeed, one can feel safe calculating concentrations of different compounds by comparing corresponding NMR signals. At the same time, the ratio of individual signals originating from the same molecules equals the ratio of the number of nuclei corresponding to these signals. For example, the measured ratio of $^1$H NMR signals in ethanol ($CH_3CH_2OH$) is precisely 3 : 2 : 1 (**Figure 2**).

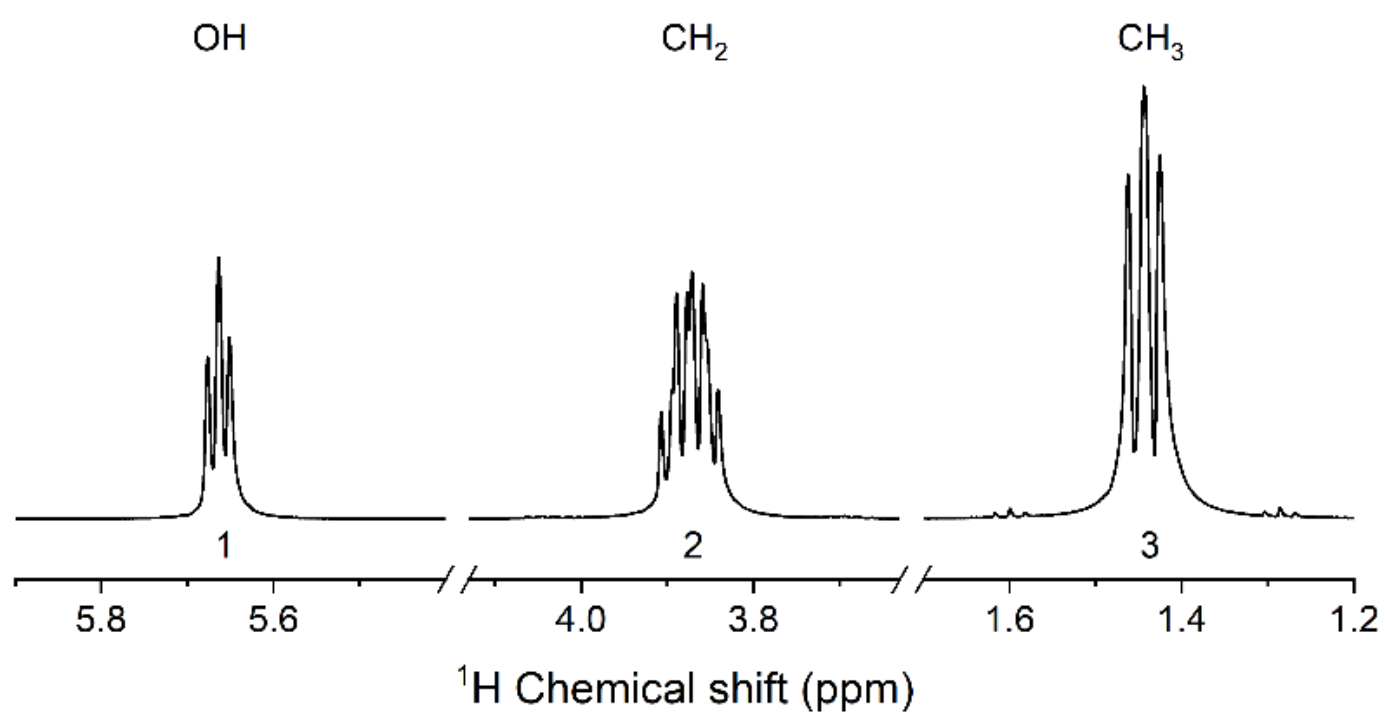

**Figure 2. $^1$H NMR spectrum of ethanol ($CH_3CH_2OH$) measured at 9.4 T and 293 K.** The implication of the theorem (eq. (18)) is that the magnetization of each chemical group (being proportional to the integral under the corresponding spectral lines) in high-field NMR spectra is linearly proportional to the number of protons irrespective of their magnetic equivalence. The sample is 600 µL of ethanol (>99.5%, 83813.360, TechniSolv) in a 5 mm NMR tube.

## Problems

Until now, we discussed magnetization of spin systems in general cases and came up with a ubiquitously used by NMR spectroscopists fact formulated as the **Theorem**. In this section, we prepared a few problems to test understanding of the mathematics behind the theorem and to derive it for some specific cases.

Problem #1. Thermal magnetization of two coupled spins-½

Compare magnetic moments of three thermally polarized samples of the same volume and geometry at high magnetic field. Sample 1 consists of $N$ non-interacting spins-½ particles. Sample 2 consists of $N/2$ pairs (molecules) of spins-½. Sample 3 consists of $N$ non-interacting spins-1 particles. Use eq. (8) and eq. (13) for the derivation of magnetizations $m_{\mathrm{z}}^{\{1/2\},N}$, $m_{\mathrm{z}}^{\{2\times(1/2)\},N/2}$, and $m_{\mathrm{z}}^{\{1\},N}$ without HT approximation. Express $m_{\mathrm{z}}^{\{1\},N}$ through $m_{\mathrm{z}}^{\{1/2\},N}$ for the cases $\mathbb{B} \gg 1$ and $\mathbb{B} \ll 1$.

The detailed solution is given in **Appendix 5**.

Problem #2. NMR signal intensities for coupled spin systems using Fermi's Golden rule

Find NMR signal intensities for the three samples introduced above using Fermi's Golden rule by considering weak continuous wave (CW) excitation of NMR transitions. For the Sample 2, compare total signals for AX and $A_2$ systems. Use the Zeeman basis set for the cases of the single spin-1/2 and AX pair and S-T basis for the $A_2$ system. This problem is instructive for better understanding the connection between spectral lines and NMR energy levels.

The detailed solution is given in **Appendix 6**.

Problem #3. Magnetization of a coupled spin system as if it is comprised of quasi-particles

Two equivalent spins-½ can be naturally described using S-T basis. Singlet state corresponds to a quasi-particle with spin-0 and triplet state corresponds to the quasi-particle with spin-1. Only spin-1 particles contribute to the magnetization. Therefore, instead of calculating magnetization for a coupled two spin-½ system, one could estimate magnetization of a system made of spin-1 particles.

Here, we propose to

1) Given HT approximation, calculate magnetization of *N* pairs of spins-½ considering that it is comprised of two independent quasi-particles with spin-0 and spin-1.
2) Show that this approach of combining magnetization of quasi-particles is possible also without HT approximation.

*Solution hint:*

Note that given HT approximation, number of spin-0 particles is *N*/4 and the number of spin-1 particles is 3*N*/4.

The detailed solution is given in **Appendix 7**.

## Generalized spin polarization: density matrix treatment

In the following, we will introduce an alternative formulation of the Theorem 1, now derived using the density matrix approach. This method is typically not explained in a sufficient level of detail in physical chemistry classes. Therefore, first, we present a short overview of the main properties of density matrices. Second, we show which properties of the spin system are missing when only magnetization is considered; they appear naturally in the case of the density matrix description [13].

The important quality of the density matrix is that $\mathrm{Tr}(\hat{\rho}) = 1$ which means that the probability of the system to be in all available states sums up to 100%. Assuming Boltzmann distribution of populations for nuclear spins at any temperature *T*, the thermal equilibrium density matrix of a system of an arbitrary number of spins described by the Hamiltonian $\hat{H}$ is

$$\hat{\rho}^{\mathrm{th}} = \frac{e^{-\frac{\hat{H}}{k_{\mathrm{B}}T}}}{\mathrm{Tr}\left(e^{-\frac{\hat{H}}{k_{\mathrm{B}}T}}\right)}. \quad (19)$$

This expression is very similar to the familiar Boltzmann distribution (eq. 13), but now written in the matrix form. The obvious useful equality of $\hat{\rho}^{\mathrm{th}}$ is commutation with the Hamiltonian since $\hat{H}e^{-\frac{\hat{H}}{k_{\mathrm{B}}T}} = e^{-\frac{\hat{H}}{k_{\mathrm{B}}T}}\hat{H}$. Because we only consider cases of thermodynamic equilibrium, we will omit the superscript "th" when mentioning $\hat{\rho}$ in the following.

A typical form of a nuclear spin Hamiltonian in liquid state at high fields is

$$\hat{H} \overset{\mathrm{HF}}{\rightarrow} 2\pi\hbar \textstyle\sum_{r=1}^{N} \nu_r \hat{I}_{r\mathrm{z}}, \quad (20)$$

where $2\pi\nu_r = \omega_r = -\gamma_r B$ is Larmor precession frequency and $\gamma_r$ is a gyromagnetic ratio for the $r$-th type of spin. An operator of the z-polarization of $r$-th spin is

$$\hat{P}_{\mathrm{z}}^{\{I_r\}} = \hat{I}_{r\mathrm{z}}/I_r. \quad (21)$$

Then, polarization of the r-th type of spins (an expectation value of $\hat{P}_{\mathrm{z}}^{\{I_r\}}$) is calculated as

$$P_{\mathrm{z}}^{\{I_r\}} = \langle \hat{P}_{\mathrm{z}}^{\{I_r\}} \rangle = \mathrm{Tr}\left(\hat{I}_{r\mathrm{z}}\hat{\rho}\right)/I_r. \quad (22)$$

Some authors find it convenient to use normalized irreducible tensors [8]. For the spin operator $\hat{I}_{r\mathrm{z}}$ corresponding irreducible tensor, using eq. A2.5, is

$$\widehat{\mathbb{T}}_{10}^{\{I_r\}} = \frac{\hat{I}_{rz}}{\mathrm{Tr}(\hat{I}_{rz}\hat{I}_{rz})^{0.5}} = \frac{\hat{I}_{rz}}{\left(\sum_{I_{rz}} I_{rz}^2\right)^{0.5}} = \frac{\hat{I}_{rz}}{\left(\frac{1}{3}I_r(I_r+1)(2I_r+1)\right)^{0.5}}. \quad (23)$$

Therefore, polarization can be calculated as

$$P_{\mathrm{z}}^{\{I_r\}} = \mathrm{Tr}\left(\widehat{\mathbb{T}}_{10}^{\{I_r\}\dagger}\hat{\rho}\right)\left(\frac{1}{3I_r}(I_r+1)(2I_r+1)\right)^{0.5}. \quad (24)$$

Single spin-½.

Before going to a deeper analysis of multispin systems, let us first consider a case of independent spin-½ particles. The corresponding matrix representation of spin and polarization operators in Zeeman basis $\{|\alpha\rangle, |\beta\rangle\}$ are as follows:

$$\hat{I}_{\mathrm{z}} = \frac{1}{2}\begin{pmatrix} 1 & 0 \\ 0 & -1 \end{pmatrix}. \quad (25)$$

For spins other than ½, the matrix representation of the operator $\hat{I}_{\mathrm{z}}$ is different but, in the Zeeman basis, it is a diagonal matrix with rank (2*I*+1 which consists of possible spin projections from *I* to –*I* with the step of 1. Therefore, one can find that $\mathrm{Tr}(\hat{I}_{\mathrm{z}}) = 0$, $\mathrm{Tr}(\hat{I}_{\mathrm{z}}\hat{I}_{\mathrm{z}}) = \sum_{I_z} I_{\mathrm{z}}^2 = \frac{I(I+1)(2I+1)}{3}$; the latter equality was proved in the Appendix 2.

Now, using HF Hamiltonian (20), the density matrix given HT approximation is

$$\hat{\rho}^{\{I\},\mathrm{HT}} = \frac{\hat{1}}{2I+1} + \frac{\mathbb{B}}{2I+1}\hat{I}_{\mathrm{z}} + o(\mathbb{B}), \quad (26)$$

since $\mathrm{Tr}\left(e^{-\frac{\hat{H}}{k_{\mathrm{B}}T}}\right) \cong \mathrm{Tr}(1) = 2I+1$. Now, one can calculate the polarization of a single spin-*I:*

$$P_{\mathrm{z}}^{\{I\},\mathrm{HT}} = \frac{1}{I}\mathrm{Tr}\left(\hat{I}_{\mathrm{z}}\hat{\rho}^{\{I\},\mathrm{HT}}\right) = \frac{\mathbb{B}}{I(2I+1)}\mathrm{Tr}\left(\hat{I}_{\mathrm{z}}\hat{I}_{\mathrm{z}}\right) = \frac{I+1}{3}\mathbb{B}. \quad (27)$$

This expression coincides with the result achieved using magnetization (eq. 15). In the same way, the polarization can be found without HT approximation.

Density matrix of a multispin system

To calculate the polarization from an arbitrary size density matrix, we need to define spin operators for a spin system composed of *n* arbitrary spins $\{I_1 I_2 \ldots I_n\}$:

$$\begin{aligned} \hat{I}_{1\mathrm{z}}^{\{I_1 I_2 \ldots I_n\}} &= \hat{I}_{1\mathrm{z}} \otimes \hat{1}_2 \otimes \hat{1}_3 \otimes \hat{1}_4 \otimes \hat{1}_5 \ldots \otimes \hat{1}_n, \\ \hat{I}_{r\mathrm{z}}^{\{I_1 I_2 \ldots I_n\}} &= \hat{1}_1 \otimes \ldots \hat{1}_{\mathrm{r}-1} \otimes \hat{I}_{r\mathrm{z}} \otimes \hat{1}_{\mathrm{r}+1} \ldots \otimes \hat{1}_n, \\ \hat{I}_{n\mathrm{z}}^{\{I_1 I_2 \ldots I_n\}} &= \hat{1}_1 \otimes \hat{1}_2 \otimes \hat{1}_3 \otimes \hat{1}_4 \otimes \hat{1}_5 \ldots \otimes \hat{I}_{n\mathrm{z}}. \end{aligned} \quad (28)$$

where $\otimes$ is a direct product, $\hat{1}_r$ is an identity matrix of the size $(2I_r+1)$ and $\hat{I}_{r\mathrm{z}}$ is the spin operator of an isolated *r*-th spin. The identity matrix of this system correspondingly equals to

$$\hat{1}^{\{I_1 I_2 \ldots I_n\}} = \hat{1}_1 \otimes \hat{1}_2 \ldots \ldots \otimes \hat{1}_n. \quad (29)$$

Then, given HT approximation, the thermal equilibrium density matrix of *N* arbitrary spins is

$$\hat{\rho}^{\{I_1 I_2 \ldots I_n\},\mathrm{HT}} = \frac{\hat{1}^{\{I_1 I_2 \ldots I_n\}}}{\prod_{m=1}^{n}(2I_m+1)} + \frac{1}{\prod_{m=1}^{n}(2I_m+1)}\sum_{r=1}^{N}\mathbb{B}_r\hat{I}_{r\mathrm{z}}^{\{I_1 I_2 \ldots I_n\}} + o(\mathbb{B}). \quad (30)$$

Here, $\mathbb{B}_r = \gamma_r\hbar B/k_{\mathrm{B}}T$. From this, polarization of any spin-$I_r$ in an arbitrary coupled spin system is

$$P_{rz}^{\{I_1 I_2 \dots I_n\},\mathrm{HT}} = \frac{1}{I_r}\mathrm{Tr}\left(\hat{I}_{rz}^{\{I_1 I_2 \dots I_n\}}\hat{\rho}^{\{I_1 I_2 \dots I_n\},\mathrm{HT}}\right) = \frac{\mathbb{B}_r}{I_r \prod_{m=1}^{N}(2I_m+1)}\mathrm{Tr}\left(\hat{I}_{rz}^{\{I_1 I_2 \dots I_n\}}\hat{I}_{rz}^{\{I_1 I_2 \dots I_n\}}\right) = \frac{\mathbb{B}_\mathrm{r}}{I_r(2I_r+1)}\mathrm{Tr}\left(\hat{I}_{rz}\hat{I}_{rz}\right) = \frac{I_r+1}{3}\mathbb{B}_r. \tag{31}$$

To calculate eq. (31), we used the property of a trace operator with regard to a direct product that allows one to group corresponding matrices: $\mathrm{Tr}\left(\hat{I}_{rz}^{\{I_1 I_2 \dots I_n\}}\hat{I}_{rz}^{\{I_1 I_2 \dots I_n\}}\right) = \mathrm{Tr}\left(\hat{1}_1\hat{1}_1\right)\cdot\mathrm{Tr}\left(\hat{1}_2\hat{1}_2\right)\cdot \dots \mathrm{Tr}\left(\hat{I}_{rz}\hat{I}_{rz}\right)\cdot \dots \cdot \mathrm{Tr}\left(\hat{1}_n\hat{1}_n\right)$ and each $\mathrm{Tr}\left(\hat{1}_m\hat{1}_m\right) = \mathrm{Tr}\left(\hat{1}_m\right) = 2I_m + 1$. Hence, polarization with respect to any spin type for a multispin system is the same as for a system of isolated non-interacting spins. This leads to a second and equivalent formulation of **Theorem 1**, which is now proven with the density matrices.

### Theorem 1 (alternative formulation)

Polarization of a spin-$I_r$ in any spin system at thermal equilibrium and at a high field ($\gamma\hbar B_\mathrm{z} \gg \|H_{\mathrm{spin-spin}}\|$) equals the polarization of an isolated spin-$I_r$ at the same magnetic field

$$P_{rz}^{\{I_1 I_2 \dots I_n\}} = P_\mathrm{z}^{\{I_r\}}. \tag{32}$$

Considering that $\langle\mu_\mathrm{z}^{I_r}\rangle = (\gamma\hbar I_r)\cdot P_\mathrm{z}^{\{I_r\}}$, and the additive nature of magnetization, the equivalence of two formulations is apparent. Eq. (31) proves the eq. (32) only at HT approximation, while the general case is considered in **Appendix 8**.

### Longitudinal two-spin order in a two-spin-½ system

So far, we have considered polarization of a spin. However, in multispin systems, higher orders are also populated. This section will consider a two-spin-½ system and find its equilibrium state with higher precision than in eq. 30. Molecular $H_2$ or methylene groups of various carbohydrates or amino acids represent typical two-spin-½ systems. The density matrix of two spins-½ up to the second order ($o(\mathbb{B}^2)$) is:

$$\hat{\rho}^{\{\frac{1}{2}\frac{1}{2}\},\mathrm{HT}} = \frac{e^{\mathbb{B}\hat{I}_{1z}^{\{\frac{1}{2}\frac{1}{2}\}}}e^{\mathbb{B}\hat{I}_{2z}^{\{\frac{1}{2}\frac{1}{2}\}}}}{\mathrm{Tr}\left(e^{\mathbb{B}\hat{I}_{1z}^{\{\frac{1}{2}\frac{1}{2}\}}}e^{\mathbb{B}\hat{I}_{2z}^{\{\frac{1}{2}\frac{1}{2}\}}}\right)} \cong \frac{\left(\hat{1}+\mathbb{B}\hat{I}_{1z}^{\{\frac{1}{2}\frac{1}{2}\}}+\frac{\mathbb{B}^2}{8}\hat{1}\right)\left(\hat{1}+\mathbb{B}\hat{I}_{2z}^{\{\frac{1}{2}\frac{1}{2}\}}+\frac{\mathbb{B}^2}{8}\hat{1}\right)}{4\left(1+\frac{\mathbb{B}^2}{4}\right)} = \frac{\hat{1}}{4} + \frac{\mathbb{B}}{4}\left(\hat{I}_{1z}^{\{\frac{1}{2}\frac{1}{2}\}} + \hat{I}_{2z}^{\{\frac{1}{2}\frac{1}{2}\}}\right) + \frac{\mathbb{B}^2}{4}\hat{I}_{1z}^{\{\frac{1}{2}\frac{1}{2}\}}\hat{I}_{2z}^{\{\frac{1}{2}\frac{1}{2}\}} + o(\mathbb{B}^2). \tag{33}$$

The maximum value of $\langle\hat{I}_{1z}^{\{\frac{1}{2}\frac{1}{2}\}}\hat{I}_{2z}^{\{\frac{1}{2}\frac{1}{2}\}}\rangle$ equals to ¼, while in general, it is equal to the product of maximum projections of two spins, $I_1 I_2$. Hence, it is reasonable to define polarization of such two-spin order as

$$P_{rtzz}^{\{I_1 I_2 \dots I_n\}} = \frac{\langle\hat{I}_{rz}^{\{I_1 I_2 \dots I_n\}}\hat{I}_{tz}^{\{I_1 I_2 \dots I_n\}}\rangle}{I_r I_t} \tag{34}$$

for any spins $I_r$ and $I_t$ in a multi-spin system. Using eqs. (33) and (34) we can now calculate polarization of such two-spin polarization under HT approximation:

$$P_{12zz}^{\{\frac{1}{2}\frac{1}{2}\},\mathrm{HT}} = \frac{\langle\hat{I}_{1z}^{\{\frac{1}{2}\frac{1}{2}\}}\hat{I}_{2z}^{\{\frac{1}{2}\frac{1}{2}\}}\rangle}{I_1 I_2} = \frac{\mathrm{Tr}\left(\hat{I}_{1z}^{\{\frac{1}{2}\frac{1}{2}\}}\hat{I}_{2z}^{\{\frac{1}{2}\frac{1}{2}\}}\hat{I}_{1z}^{\{\frac{1}{2}\frac{1}{2}\}}\hat{I}_{2z}^{\{\frac{1}{2}\frac{1}{2}\}}\right)}{0.25}\frac{\mathbb{B}^2}{4} = \mathrm{Tr}\left(\hat{I}_\mathrm{z}\hat{I}_\mathrm{z}\otimes\hat{I}_\mathrm{z}\hat{I}_\mathrm{z}\right)\mathbb{B}^2 = = \left[\mathrm{Tr}\left(\hat{I}_\mathrm{z}\hat{I}_\mathrm{z}\right)\right]^2\mathbb{B}^2 = \frac{\mathbb{B}^2}{4} = \left(P_\mathrm{z}^{\{1/2\},\mathrm{HT}}\right)^2. \tag{35}$$

Above, we omitted spin indices after the third equality because only spin-½ particles are discussed and they have the same spin operators (eq. 25) and $\mathrm{Tr}\left(\hat{I}_\mathrm{z}\hat{I}_\mathrm{z}\right) = \frac{1}{2}$.

Without HT approximation, it can be calculated that

$$P_{12\mathrm{zz}}^{\{½½\}} = \frac{\mathrm{Tr}\left(\hat{I}_{1\mathrm{z}}^{\{½½\}}\hat{I}_{2\mathrm{z}}^{\{½½\}} e^{\mathbb{B}\hat{I}_{1\mathrm{z}}^{\{½½\}}} e^{\mathbb{B}\hat{I}_{2\mathrm{z}}^{\{½½\}}}\right)}{I_1 I_2 \cdot \mathrm{Tr}\left(e^{\mathbb{B}\hat{I}_{1\mathrm{z}}^{\{½½\}}} e^{\mathbb{B}\hat{I}_{2\mathrm{z}}^{\{½½\}}}\right)} = \left(\frac{2\mathrm{Tr}\left(\hat{I}_\mathrm{z} e^{\mathbb{B}\hat{I}_\mathrm{z}}\right)}{\mathrm{Tr}\left(e^{\mathbb{B}\hat{I}_\mathrm{z}}\right)}\right)^2 = \left(P_\mathrm{z}^{\{½\}}\right)^2. \tag{36}$$

The first and the third equalities are definitions of polarization, while the second equality is trivially derived using the same idea as in eq. (35): different spins commute and the trace of direct products is the product of tracers. Hence, at thermal equilibrium, $P_{12\mathrm{zz}}^{\{½½\}} = \left(P_\mathrm{z}^{\{½\}}\right)^2$ and combining with eq. (14) it gives a functional dependence of $P_{12\mathrm{zz}}^{\{½½\}}$ on $\mathbb{B}$. Both values, $P_{12\mathrm{zz}}^{\{½½\}}$ and $P_\mathrm{z}^{\{½\}}$, asymptotically reach unity but the two-spin order is reaching it "slower" (**Figure 3**).

### Density operator representation for multispin systems

Any operator can be decomposed in terms of its basis states. Irreducible tensors are very common as such a basis in NMR because they are naturally orthonormal. The other approach is based on product operators since they are easy to visualize, and their evolution can be described geometrically. Following eq. (33) and definitions of $P_\mathrm{z}^{\{½\}}$ and $P_{12\mathrm{zz}}^{\{½½\}}$, the density matrix of two spin-½ systems can be derived:

$$\hat{\rho}^{\{½½\}} = \frac{\hat{1}^{\{½½\}}}{4} + \frac{1}{2}P_{1\mathrm{z}}^{\{½\}} \cdot \hat{I}_{1\mathrm{z}}^{\{½½\}} + \frac{1}{2}P_{2\mathrm{z}}^{\{½\}} \cdot \hat{I}_{2\mathrm{z}}^{\{½½\}} + P_{12\mathrm{zz}}^{\{½½\}} \cdot \hat{I}_{1\mathrm{z}}^{\{½½\}}\hat{I}_{2\mathrm{z}}^{\{½½\}}. \tag{37}$$

This equation gives the density matrix of the system with a polarization of $P_{1\mathrm{z}}^{\{½\}}$ and $P_{2\mathrm{z}}^{\{½\}}$ and two-spin order polarization $P_{12\mathrm{zz}}^{\{½½\}}$. When thermal equilibrium polarization values are used, the resulting matrix is also at thermal equilibrium.

Analogously, one can obtain that for *N* spin-½ the density matrix can be then written in the following way:

$$\hat{\rho}^{\{1/2\otimes N\}} = \frac{\hat{1}^{\{1/2\otimes N\}}}{2^N} + \frac{1}{2^{N-1}}\sum_{r=1}^{N} P_{r\mathrm{z}}^{\{½\}} \cdot \hat{I}_{r\mathrm{z}}^{\{1/2\otimes N\}} + \frac{1}{2^{N-2}}\sum_{1\le r<m\le N} P_{rm\mathrm{zz}}^{\{½½\}} \cdot \hat{I}_{r\mathrm{z}}^{\{1/2\otimes N\}}\hat{I}_{m\mathrm{z}}^{\{1/2\otimes N\}} + \cdots. \tag{38}$$

And finally for *N* arbitrary spins, the density matrix is given by the following expression:

$$\hat{\rho}^{\{I_1 I_2 \dots I_n\}} = \frac{\hat{1}^{\{I_1 I_2 \dots I_n\}}}{\mathbb{Z}} + \frac{1}{\mathbb{Z}}\sum_{r=1}^{N} \frac{3P_{r\mathrm{z}}^{\{I_r\}}}{I_r+1}\hat{I}_{r\mathrm{z}}^{\{I_1 I_2 \dots I_n\}} + \frac{1}{\mathbb{Z}}\sum_{1\le r<m\le N} \frac{9P_{rm\mathrm{zz}}^{\{I_r I_\mathrm{m}\}}}{(I_r+1)(I_m+1)}\hat{I}_{r\mathrm{z}}^{\{I_1 I_2 \dots I_n\}}\hat{I}_{m\mathrm{z}}^{\{I_1 I_2 \dots I_n\}} + \cdots. \tag{39}$$

where $\mathbb{Z} = \prod_{r=1}^{N}(2I_r + 1)$. The polarization of other spin orders can be derived and added in the same way if necessary. In the following section, we will derive one more term in the density matrix, which is present for spin-*I* particles with *I* > ½ and which so far has not been considered by us.

### Quadrupolar polarization of a system with spin-*I* (*I*>½)

In the previous section, we discussed how different spin orders can be populated in multispin systems. Here we will show that even in a single spin-*I* system, a state could not be fully described with polarization alone. Specifically, eq. (26) was obtained up to the first order of precision for the magnetic field, $o(\mathbb{B})$. Let us have a look at the same density matrix of a single spin-*I* with a higher precision using HT approximation:

$$\hat{\rho}^{\{I\},\mathrm{HT}} = \frac{e^{\mathbb{B}\hat{I}_\mathrm{z}}}{\mathrm{Tr}\left(e^{\mathbb{B}\hat{I}_\mathrm{z}}\right)} \cong \frac{\hat{1}+\mathbb{B}\hat{I}_\mathrm{z}+\frac{\mathbb{B}^2}{2}(\hat{I}_\mathrm{z})^2}{\mathrm{Tr}\left(\hat{1}+\mathbb{B}\hat{I}_\mathrm{z}+\frac{\mathbb{B}^2}{2}(\hat{I}_\mathrm{z})^2\right)} = \frac{\hat{1}}{2I+1} + \frac{\mathbb{B}}{2I+1}\hat{I}_\mathrm{z} + \frac{1}{2I+1}\frac{\mathbb{B}^2}{6}\left[3\left(\hat{I}_\mathrm{z}\right)^2 - I(I+1)\hat{1}\right] + o(\mathbb{B}^2). \tag{40}$$

Here, the denominator was simplified as $\mathrm{Tr}\left(\hat{1} + \mathbb{B}\hat{I}_\mathrm{z} + \frac{\mathbb{B}^2}{2}\left(\hat{I}_\mathrm{z}\right)^2\right) = \mathrm{Tr}\left(\hat{1} + \frac{\mathbb{B}^2}{2}\left(\hat{I}_\mathrm{z}\right)^2\right) = (2I+1) + \frac{\mathbb{B}^2}{2}\sum_{I_\mathrm{z}=-I}^{I}(I_\mathrm{z})^2 = (2I+1)\left(1 + \frac{\mathbb{B}^2}{2}\frac{I(I+1)}{3}\right)$.

This expansion looks very similar to the multipole expansion used, for example, in electrodynamics with dipolar ($\hat{I}_z$) and quadrupolar ($3\left(\hat{I}_z\right)^2 - I(I+1)\hat{1}$) magnetic moments; the latter term is also often referred to as the "alignment" [12]. Another important fact is that for spin-½ systems $\left(\hat{I}_z\right)^2 = \hat{1}/4$ and, hence, quadrupolar moment equals to zero. One can think of spin-½ as of purely dipolar nature. In all other cases, the trace of the enclosed in the square brackets operator in eq. (40) is nonzero.

Following the previous normalization concept, the operator of quadrupolar polarization is as follows:

$$\hat{P}_{\mathrm{Q}}^{\{I\}} = \frac{3(\hat{I}_z)^2 - I(I+1)\hat{1}}{I^2}. \quad (41)$$

The corresponding quadrupole polarization (alignment) is (see derivation in **Appendix 9**):

$$P_{\mathrm{Q}}^{\{I\}} = \frac{2(I+1)}{I} + \frac{3}{2I} + \frac{3}{2\sinh^2\left(\frac{\mathbb{B}}{2}\right)} - \frac{3}{I^2}\left(I+\frac{1}{2}\right)\coth\left(\left(I+\frac{1}{2}\right)\mathbb{B}\right)\coth\left(\frac{\mathbb{B}}{2}\right). \quad (42)$$

Unlike $P_{\mathrm{z}}^{\{I\}}$, $P_{\mathrm{Q}}^{\{I\}}$ does not change from -1 to +1. For example, for spin-1, $P_{\mathrm{Q}}^{\{I\}}$ can change from -2 to 1. However, at the thermal equilibrium $P_{\mathrm{Q}}^{\{I\}} \in [0,1]$ disregarding the sign of the gyromagnetic ratio **(Figure 3)**.

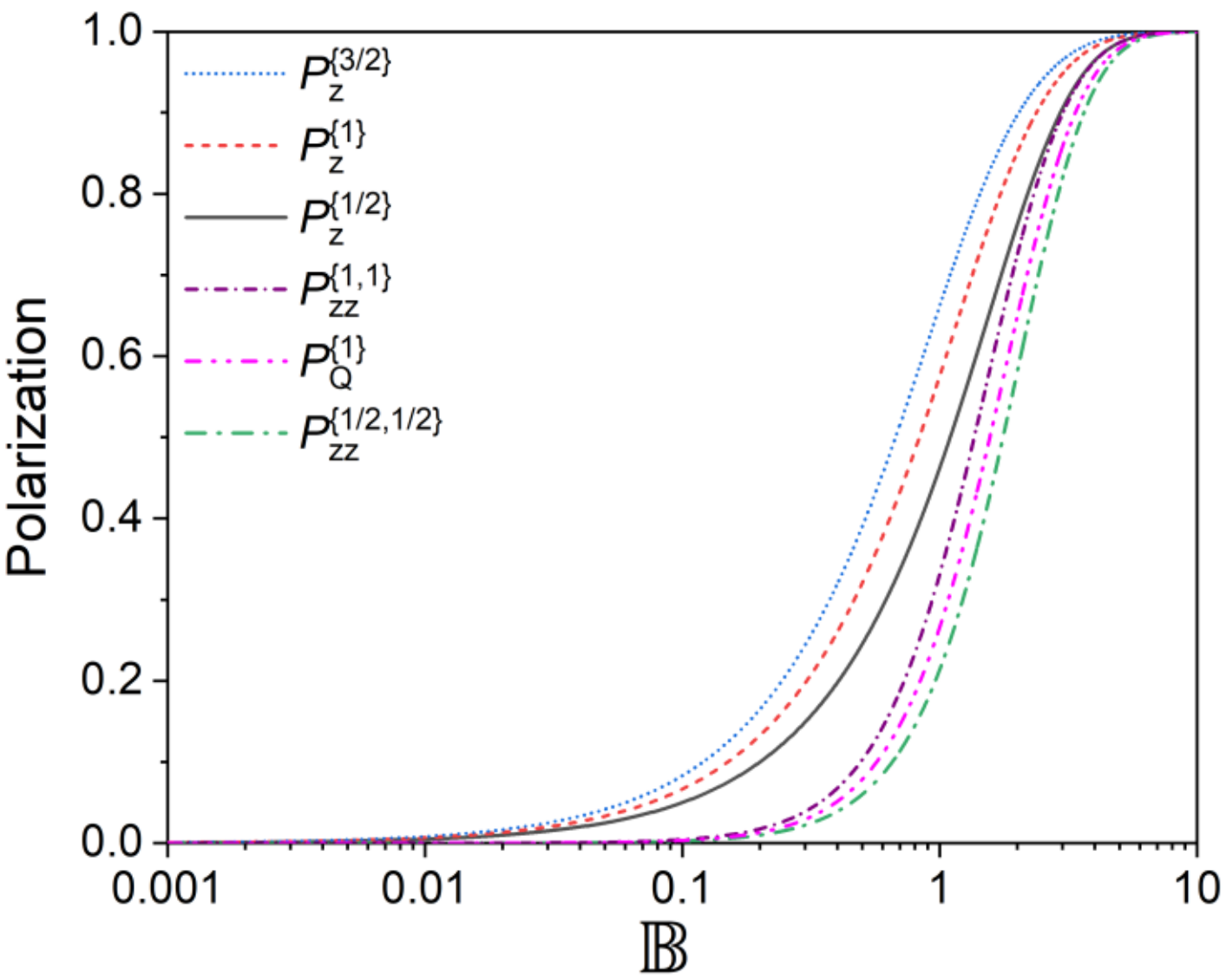

**Figure 3.** Polarization $P_{\mathrm{z}}^{\{I\}}$ for spins-½, 1, and 3/2, alignment $P_{\mathrm{Q}}^{\{I\}}$ for spin-1 and $P_{\mathrm{zz}}^{\{I,I\}}$ for pairs of spin-½ and spin-1 systems as a function of $\mathbb{B}$ (Boltzmann factor).

## Is pure 100% polarization possible?

Above, we found the density matrices for two-spin-½ and multispin systems which comprise single-spin polarizations, quadrupolar alignment, and multispin order. For example, for two-spins-½, the system in equilibrium is characterized by three values: $P_{1\mathrm{z}}^{\{½\}}$, $P_{2\mathrm{z}}^{\{½\}}$, and $P_{12\mathrm{zz}}^{\{½½\}}$ each is in the range from -1 to +1. If all the values were possible, then the volume of the configuration space would have been 8 (a cube with a side of 2). Below, we will demonstrate that there are constrains on possible polarization values that follow from the fundamental properties of density matrices.

By definition, (i) the density matrix is normalized, $\mathrm{Tr}(\hat{\rho}) = 1$, meaning that the probability of finding the system in any state is 1, (ii) it is also a Hermitian operator, $\hat{\rho} = \hat{\rho}^{\dagger}$, and (iii) diagonal elements are non-negative and do not exceed 1, $1 \geq \rho_r = \rho_{rr} \geq 0$ (meaning that the probability of finding the system in any state is in the range of 0 to 1).

The first two requirements are immediately fulfilled for all the density matrices given above because all operators but $\hat{1}$ are traceless and Hermitian. However, the last requirement puts unexpected restrictions. Considering $1 \geq \rho_r = \rho_{rr} \geq 0$, the allowed volume for the spin system shrinks from 8 to 4/3 (**Figure 4**, derived in **Appendix 10**). One should note that in addition to restrictions on populations, there are symmetry restrictions on spin order transfer depending on the spin topology that we do not discuss here [14,15].

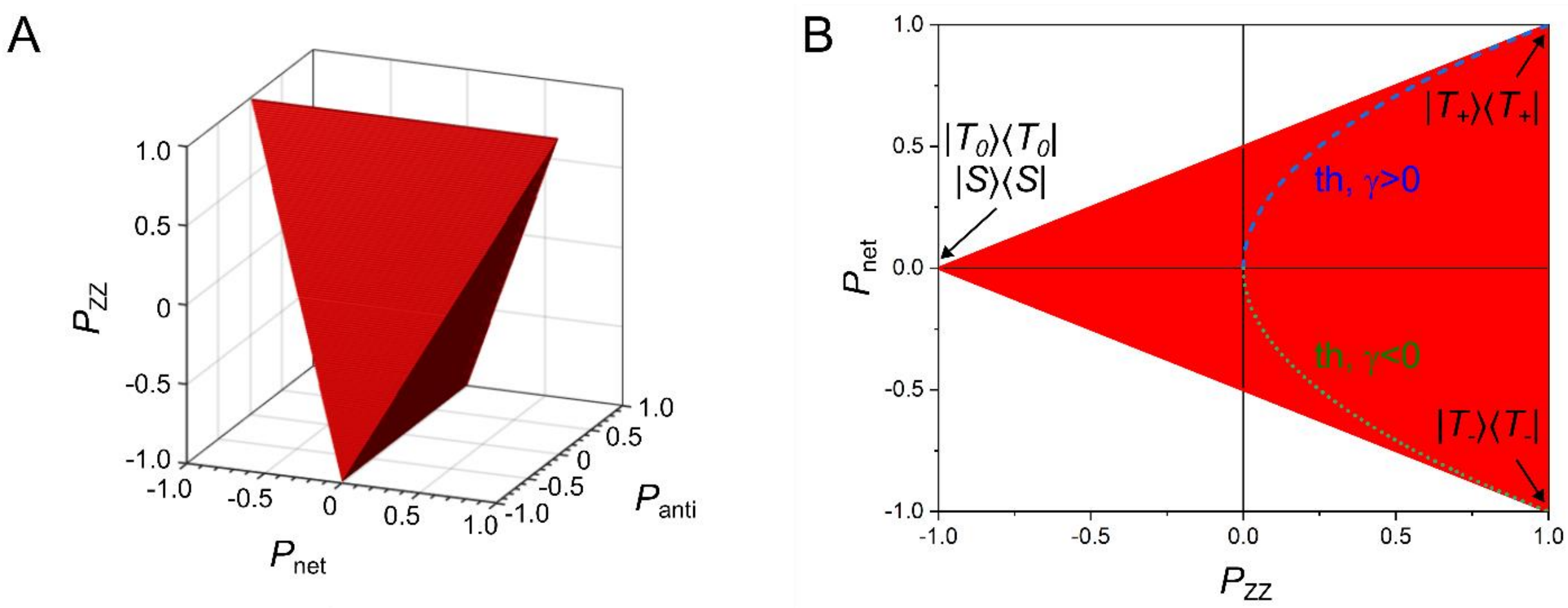

**Figure 4.** (A) The volume of allowed values of polarization for density matrices in units of $P_{\mathrm{net}} = \left(P_{1\mathrm{z}}^{\{½\}} + P_{2\mathrm{z}}^{\{½\}}\right)/2$, $P_{\mathrm{anti}} = \left(P_{1\mathrm{z}}^{\{½\}} - P_{2\mathrm{z}}^{\{½\}}\right)/2$ and $P_{\mathrm{zz}} = P_{12\mathrm{zz}}^{\{½½\}}$ and (B) the area of allowed $P_{\mathrm{net}}$ and $P_{\mathrm{zz}}$ polarization values for $P_{\mathrm{anti}} = 0$. The dashed (blue) and dotted (green) lines indicate polarization values at thermal equilibrium for positive and negative values of the gyromagnetic ratio. The volume of allowed states in (A) is 4/3, while the nominal available volume is 8. The area of allowed states in (B) is 2, while the nominal available area is 4.

To understand how it happens in a simple example, let us find the density matrix for two spins-½ at the state with 100% polarization. The two-spin order has to be inevitably populated. To rationalize this observation, let us find the spin order representation for the 100% net magnetization. One of the ways is to write density matrices with 100% population of one of the states in the S-T basis via product operators [16]:

$$\hat{\rho}_{T_+}^{\{½½\}} = |T_+\rangle\langle T_+| = \frac{\hat{1}^{\{½½\}}}{4} + \frac{1}{2}\left(\hat{I}_{1\mathrm{z}}^{\{½½\}} + \hat{I}_{2\mathrm{z}}^{\{½½\}}\right) + \hat{I}_{1\mathrm{z}}^{\{½½\}}\hat{I}_{2\mathrm{z}}^{\{½½\}},$$
$$\hat{\rho}_{T_0}^{\{½½\}} = |T_0\rangle\langle T_0| = \frac{\hat{1}^{\{½½\}}}{4} + \left(\hat{\mathbf{I}}_1^{\{½½\}} \cdot \hat{\mathbf{I}}_2^{\{½½\}}\right) - 2\hat{I}_{1\mathrm{z}}^{\{½½\}}\hat{I}_{2\mathrm{z}}^{\{½½\}},$$
$$\hat{\rho}_{T_-}^{\{½½\}} = |T_-\rangle\langle T_-| = \frac{\hat{1}^{\{½½\}}}{4} - \frac{1}{2}\left(\hat{I}_{1\mathrm{z}}^{\{½½\}} + \hat{I}_{2\mathrm{z}}^{\{½½\}}\right) + \hat{I}_{1\mathrm{z}}^{\{½½\}}\hat{I}_{2\mathrm{z}}^{\{½½\}},$$
$$\hat{\rho}_{S}^{\{½½\}} = |S\rangle\langle S| = \frac{\hat{1}^{\{½½\}}}{4} - \left(\hat{\mathbf{I}}_1^{\{½½\}} \cdot \hat{\mathbf{I}}_2^{\{½½\}}\right), \quad (43)$$

where $\left(\hat{\mathbf{I}}_1^{\{½½\}} \cdot \hat{\mathbf{I}}_2^{\{½½\}}\right) = \hat{I}_{1x}^{\{½½\}}\hat{I}_{2x}^{\{½½\}} + \hat{I}_{1y}^{\{½½\}}\hat{I}_{2y}^{\{½½\}} + \hat{I}_{1z}^{\{½½\}}\hat{I}_{2z}^{\{½½\}}$. All these states have $P_{\mathrm{anti}} = (P_{1z}^{½} - P_{2z}^{½})/2 = 0$; $\hat{\rho}_S$ and $\hat{\rho}_{T_0}$ have $P_{\mathrm{net}} = (P_{1z}^{½} + P_{2z}^{½})/2 = 0$ and $P_{zz} = -1$; $\hat{\rho}_{T_+}$ has $P_{\mathrm{net}} = 1$ and $P_{zz} = 1$; $\hat{\rho}_{T_-}$ has $P_{\mathrm{net}} = -1$ and $P_{zz} = 1$. The last two states are in the corners of the allowed space for $P_{zz} = 1$ (**Figure 4B**).

The density matrix of the system with *only* 100% polarization ($P_{\mathrm{net}} = 1$, $P_{\mathrm{anti}} = 0$, $P_{zz} = 0$) would be as follows:

$$\hat{\rho}_{100\%}^{\{½½\}} = \frac{\hat{1}}{4} + \frac{1}{2}\left(\hat{I}_{1z}^{\{½½\}} + \hat{I}_{2z}^{\{½½\}}\right) = \frac{3|T_+\rangle\langle T_+| + |T_0\rangle\langle T_0| + |S\rangle\langle S| - |T_-\rangle\langle T_-|}{4}. \tag{44}$$

It is clear that such a state would require a negative population of $|T_-\rangle$ which is not feasible. However, the ±100% polarization is possible and given by $\hat{\rho}_{T_\pm}$ (eq. (43)) where both $|P_{\mathrm{net}}|$ and $P_{zz}$ are maximized.

From the **Figure 4B**, it follows that the highest achievable $P_{\mathrm{net}}$ given $P_{\mathrm{anti}} = P_{zz} = 0$, is only 50%: $\hat{\rho}_{50\%}^{\{½½\}} = \frac{\hat{1}}{4} + \frac{1}{4}\left(\hat{I}_{1z}^{\{½½\}} + \hat{I}_{2z}^{\{½½\}}\right) = \frac{2|T_+\rangle\langle T_+| + |T_0\rangle\langle T_0| + |S\rangle\langle S|}{4}$. No polarization ($P_{\mathrm{net}} = 0$) is allowed for $P_{zz} = -1$.

## Special case: Polarization of $H_2$ molecules

Here we would like to discuss one peculiar case where molecular degrees of freedom can have an effect on nuclear spin populations using an example of molecular hydrogen. Hydrogen ($H_2$) exists under normal conditions in two spin isomeric forms called parahydrogen ($p$H$_2$) and orthohydrogen ($o$H$_2$). Parahydrogen has a total nuclear spin-0 while $o$H$_2$ has a total spin-1 resulting in the degeneracy of 1 and 3, respectively. At room temperature conditions, the ortho-para ratio is close to 3. The ratio at an arbitrary temperature can be calculated using Boltzmann distribution for the rotational levels characterized by a rotational quantum number $J$ (magnetic field $B$=0 as indicated by superscript):

$$\frac{n_{p\mathrm{H}_2}^{0,T}}{n_{o\mathrm{H}_2}^{0,T}} = \frac{\sum_{J=2n}(2J+1)\exp(-J(J+1)\theta_{\mathrm{R}}/T)}{3\sum_{J=2n+1}(2J+1)\exp(-J(J+1)\theta_{\mathrm{R}}/T)}. \tag{45}$$

Here $E_{J=1} - E_{J=0} = 2\theta_{\mathrm{R}}$ with $\theta_{\mathrm{R}} \cong 87.6$ K [17].

The singlet-triplet transitions are forbidden for $H_2$ under normal conditions. Paramagnetic materials such as charcoal, iron (III) oxide [18,19], or metal-organic frameworks (MOFs) [20] can convert $p$H$_2$ to $o$H$_2$. Let us now find the thermal equilibrium polarization of $H_2$ for a general case in the presence of magnetic field. Since $n_S^{B,T} = n_S^{0,T} = n_{p\mathrm{H}_2}^{0,T}$, $n_{T_\pm}^{B,T} = n_{T_\pm}^{0,T} e^{\pm\mathbb{B}}$, $n_{T_0}^{B,T} = n_{T_0}^{0,T}$ and $n_{T_{+,0,-}}^{0,T} = \frac{n_{o\mathrm{H}_2}^{0,T}}{3}$, then

$$P_z^{\{\mathrm{H}_2\}} = \frac{n_{T_+}^{B,T} - n_{T_-}^{B,T}}{n_{T_+}^{B,T} + n_{T_0}^{B,T} + n_{T_-}^{B,T} + n_S^{B,T}} = \frac{e^{\mathbb{B}} - e^{-\mathbb{B}}}{e^{\mathbb{B}} + 1 + e^{-\mathbb{B}} + 3\frac{n_{p\mathrm{H}_2}^{0,T}}{n_{o\mathrm{H}_2}^{0,T}}}. \tag{46}$$

For $\mathbb{B} \ll 1$, eq. (46) can be simplified to

$$P_z^{\{\mathrm{H}_2\}} \cong \frac{2\mathbb{B}}{3\left(1 + \frac{n_{p\mathrm{H}_2}^{0,T}}{n_{o\mathrm{H}_2}^{0,T}}\right)}, \tag{47}$$

where the fraction in denominator is given by eq. (45) which at room temperature is close to 1/3, hence:

$$P_z^{\{\mathrm{H}_2\},\mathrm{HT}} \cong \frac{1}{2}\mathbb{B} = P_z^{\{1/2\},\mathrm{HT}}. \tag{48}$$

Thus, molecular hydrogen has the same polarization as atomic hydrogen, and an ensemble hydrogen molecules at equilibrium at high field gives the same NMR signal as an ensemble of protons.

It was recently hypothesized that nuclear magnetization of molecular orthohydrogen may be a reason for the magnetism of Jovian planets [21]. We analyzed achievable hydrogen polarization levels as a function of magnetic field and temperature using the eq. (46) (**Figure 5**). One can see that it is indeed possible to achieve large (>10%) nuclear polarization at a wide range of temperatures given that the magnetic field is also large (>$10^4$ T). This is reasonable since at such a magnetic field the $|T_+\rangle$ state becomes the lowest energy level, unlike a typically encountered situation where the para-state is the lowest. One can estimate whether or not such a large field could be generated by nuclear spins themselves to further induce polarization of the surrounding spins. Consider answering the following question: What is a magnetic field at the surface of a ball consisting of fully hyperpolarized hydrogen spins? Assuming density of hydrogen $\rho$ and molar mass *M*, the field at the surface of such a ball is

$$B = \frac{\mu_0}{2\pi R^3}\left(\frac{\gamma\hbar}{2}N\right) = \frac{\mu_0\gamma\hbar\rho}{3M}. \tag{49}$$

By substituting literature values (mean hydrogen concentration on Jupiter is $\rho = 1.326$ g/mL) into the above expression one can estimate $B \approx 16$ mT irrespective of the size of the ball. Therefore, it is unlikely that the magnetic field generated by nuclear spins alone could induce substantial polarization of surrounding nuclei (self-induced nuclear ferromagnetism) without other mechanisms of field generation. However, one should note that eq. (45) is based on the assumption of thermal equilibrium which does not work on the scale of Jovian planets and we have not considered possible observable manifestations of multispin orders which could potentially be present as well.

One should also consider that in the core of Jupiter and other gas giants pressure can be so high that nuclear wavefunctions overlap (it was recently shown to happen above 60 GPa [22]). Under such conditions, one can no longer assume free molecular rotation as hydrogen becomes liquid (or even solid), and the applicability of Boltzmann statistics breaks down. Therefore, one has to use Fermi-Dirac statistics to estimate polarization which is non-trivial and lies beyond the scope of this paper [23].

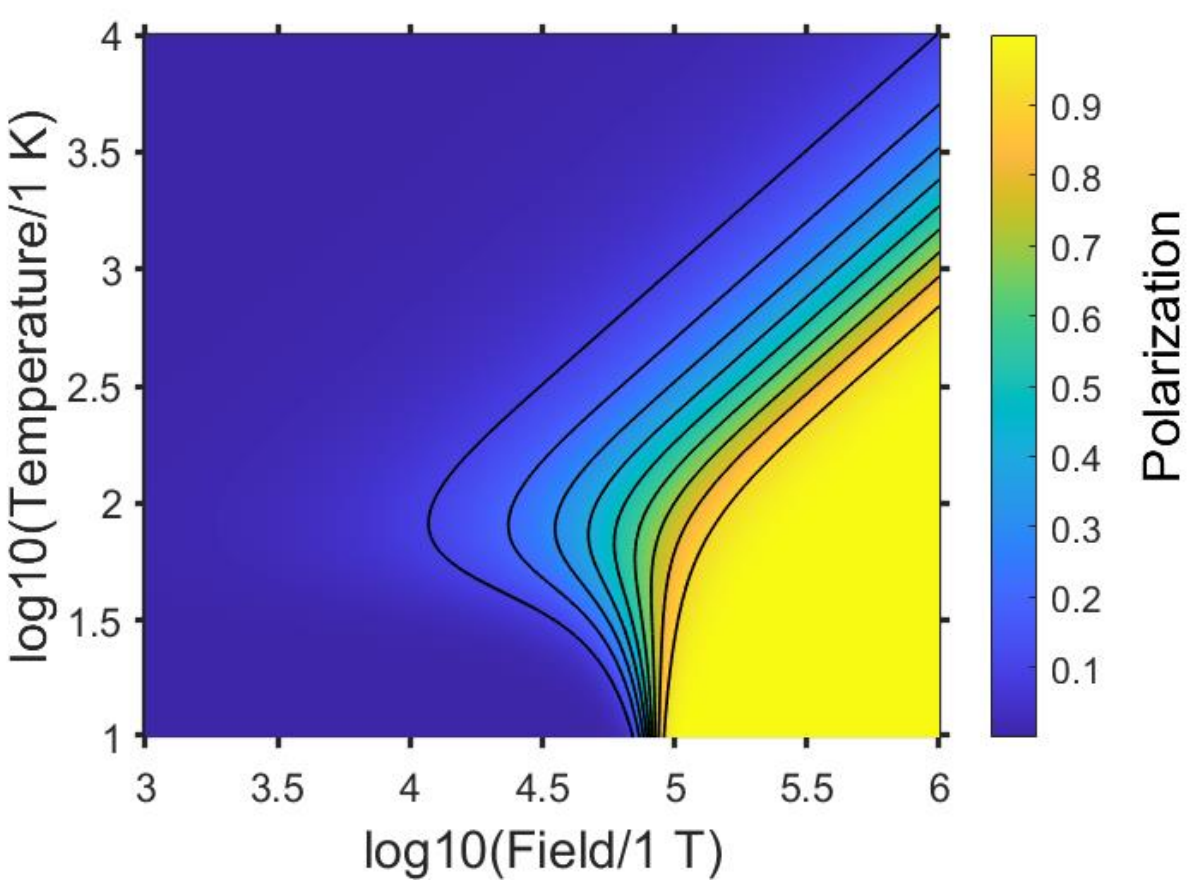

**Figure 5. Equilibrium polarization of $H_2$ as a function of temperature and magnetic field (eq. 46).** At moderate temperatures (10-100 K) and extremely high magnetic fields (~$10^5$ T) 100% polarization is feasible. Solid lines are isolines for polarization values 0.1, 0.2, 0.3, 0.4, 0.5, 0.6, 0.7, 0.8, and 0.9.

# Conclusions

In this work, we derive equations for magnetization and polarization in multispin systems of arbitrary size and spin quantum numbers, considering cases of thermodynamic equilibrium at high magnetic fields. We formulate two lemmas and a theorem, leading to two key conlusions: (i) polarization is independent of the size and topology of the spin system, and (ii) magnetization of the system of any spins under high-field conditions is an additive quantity, irrespective of weather the spins are magnetically equivalent. For educational purposes, we present two alternative formulations of the theorem---one based on classical magnetization and the other using density matrix formalism. To further support understanding, we provide three problems with detailed solutions. Lastly, we explore the space of available spin orders in a general two-spin system and discuss the feasibility of nuclear ferromagnetism in Jovian planets. The results shown here align with previous studies while opening avenues for future research into conditions where the lemmas and theorem do not hold, such as in at zero- to ultralow-magnetic fields or far from thermodynamic equilibrium.

# Acknowledgments

We thank Prof. Malcolm Levitt and Prof. Sergey Egorov for stimulating discussions. DAB acknowledges support from Alexander von Humboldt Foundation in the framework of Sofja Kovalevskaja Award. ANP acknowledges funding from German Federal Ministry of Education and Research (BMBF) within the framework of the e:Med research, funding concept (01ZX1915C), DFG (PR 1868/3-1, EXC2167). MOIN CC was funded by a grant from the European Regional Development Fund (ERDF) and the Zukunftsprogramm Wirtschaft of Schleswig-Holstein (project no. 122-09-053).

# Appendix

### A1. Thermal polarization beyond spin-½

Let us first calculate the sum in the denominator of eq. (13) which is the known sum of a geometric progression:

$$\sum_{I_z=-I}^{I} e^{\mathbb{B}I_z} = \frac{e^{(I+\frac{1}{2})\mathbb{B}} - e^{-(I+\frac{1}{2})\mathbb{B}}}{e^{\mathbb{B}/2} - e^{-\mathbb{B}/2}}. \qquad \text{(A1.1)}$$

By differentiating both parts of the above equation over the parameter $\mathbb{B}$, one obtains a numerator:

$$\frac{d}{d\mathbb{B}}\sum_{I_z=I}^{-I} e^{\mathbb{B}I_z} = \sum_{I_z=I}^{-I} I_z e^{\mathbb{B}I_z} = \left(I+\frac{1}{2}\right)\frac{e^{\left(I+\frac{1}{2}\right)\mathbb{B}} + e^{-\left(I+\frac{1}{2}\right)\mathbb{B}}}{e^{\mathbb{B}/2} - e^{-\mathbb{B}/2}} - \frac{1}{2}\left(\frac{e^{\mathbb{B}/2} + e^{-\mathbb{B}/2}}{e^{\mathbb{B}/2} - e^{-\mathbb{B}/2}}\right)\frac{e^{\left(I+\frac{1}{2}\right)\mathbb{B}} - e^{-\left(I+\frac{1}{2}\right)\mathbb{B}}}{e^{\mathbb{B}/2} - e^{-\mathbb{B}/2}}. \qquad \text{(A1.2)}$$

Therefore, by finding a ratio of (A1.2) and (A1.1) one obtains

$$\frac{\sum_{I_z=I}^{-I} I_z e^{\mathbb{B} I_z}}{\sum_{I_z=-I}^{I} e^{\mathbb{B} I_z}} = \left(I+\frac{1}{2}\right)\frac{e^{\left(I+\frac{1}{2}\right)\mathbb{B}}+e^{-\left(I+\frac{1}{2}\right)\mathbb{B}}}{e^{\left(I+\frac{1}{2}\right)\mathbb{B}}-e^{-\left(I+\frac{1}{2}\right)\mathbb{B}}} - \frac{1}{2}\left(\frac{e^{\frac{\mathbb{B}}{2}}+e^{-\frac{\mathbb{B}}{2}}}{e^{\frac{\mathbb{B}}{2}}-e^{-\frac{\mathbb{B}}{2}}}\right). \quad \text{(A1.3)}$$

From here, using the hyperbolic cotangent defined as $\coth x = (e^x + e^{-x})/(e^x - e^{-x})$ and dividing the result by $I$, polarization of a spin-*I* (eq. 14) is obtained. Using Taylor's expansion of hyperbolic cotangent, $\coth x = \frac{1}{x} + \frac{x}{3} - \cdots$, it is straightforward to show that under HT conditions ($\mathbb{B} \ll 1$, $x \ll 1$), indeed, $P_z^{\{I\}} \xrightarrow{\mathbb{B}\to 0} P_z^{\{I\},\mathrm{HT}}$ (eq. 15).

## A2. A sum of squared half-integer numbers is a sum of squared spin projections

The sum of the first *N* positive integers is known (sum of arithmetic progression with step 1):

$$\sum_{n=1}^{N} n = \frac{N(N+1)}{2}, \quad \text{(A2.1)}$$

as well as the sum of the squares of the first *N* positive integers:

$$\sum_{n=1}^{N} n^2 = \frac{N(N+1)(2N+1)}{6}. \quad \text{(A2.2)}$$

Therefore, for integer spins, the sum of squares for all possible projections $I_\mathrm{Z}$ (taking values from $-I$ to $+I$ with the step of 1) is twice this value:

$$\sum_{I_z=\mathrm{I}:1:-\mathrm{I}} I_z^2 = \frac{I(I+1)(2I+1)}{3}. \quad \text{(A2.3)}$$

The sum of the half-integer *I*-s can be found as follows. First, we find the sum of squares for *N* odd positive integers:

$$\sum_{n=1}^{N}(2n-1)^2 = 4\sum_{n=1}^{N} n^2 - 4\sum_{n=1}^{N} n + N = 2\frac{N(N+1)(2N+1)}{3} - 2N(N+1) + N = \\ = \frac{N(2N+1)(2N-1)}{3}. \quad \text{(A2.4)}$$

If we now divide this expression by 4 and take it twice, we will obtain the requested sum of squares for half-integer numbers. The maximum spin projection *I* corresponds here to $\frac{2N-1}{2}$ and minimum to ½. After substituting $N \to I + 1/2$, one obtains

$$2\sum_{n=1}^{N}\frac{(2n-1)^2}{4} = \frac{N(2N+1)(2N-1)}{6} \to \frac{\left(I+\frac{1}{2}\right)(2I+2)(2I)}{6} = \frac{I(I+1)(2I+1)}{3}. \quad \text{(A2.5)}$$

Hence eq. A2.3 is correct for both integer and half-integer spin quantum numbers.

## A3. Sums involving binomial coefficients

Different sums of binominal coefficients appear when one explicitly calculates magnetization of an arbitrary number of spin-½ particles using HT approximation. Here, we will find some sums that will be later used for the derivations of magnetization values. The following three sums involving binomial coefficients $\binom{n}{k} = C_n^k$ are well-known:

$$\sum_{0}^{n} C_n^k = 2^n, \quad \text{(A3.1)}$$

$$\sum_{k=0}^{n} k C_n^k = n 2^{n-1}, \quad \text{(A3.2)}$$

$$\sum_{k=0}^{n} k^2 C_n^k = (n+n^2)2^{n-2}. \quad \text{(A3.3)}$$

Therefore, the quantity $\sum_{k=0}^{n} C_n^k \left(\frac{n}{2}-k\right)^2$ can be found using (A3.1-3):

$$\sum_{k=0}^{n} C_n^k \left(\frac{n}{2}-k\right)^2 = \sum_0^n C_n^k \left(\frac{1}{4}n^2 - nk + k^2\right) = \frac{1}{4}n^2 \sum_0^n C_n^k - n\sum_{k=0}^{n} kC_n^k + \sum_{k=0}^{n} k^2 C_n^k = \tag{A3.4}$$
$$= \frac{1}{4}n^2 2^n - \frac{1}{2}n^2 2^n + \frac{1}{4}(n+n^2)2^n = \frac{1}{4}n2^n.$$

Another useful sum involves exponentials encountered when computing magnetization of an arbitrary number of spins-½ without using HT approximation. The following equality follows from the definition of binomial coefficients:

$$\left(e^{\frac{\mathbb{B}}{2}} + e^{-\frac{\mathbb{B}}{2}}\right)^n = \sum_{k=0}^{n} C_n^k e^{\left(\frac{n}{2}-k\right)\mathbb{B}}, \tag{A3.5}$$

The right part of this equation is a denominator when such magnetization is calculated.

Derivative of the left part of the eq. (A3.5) with respect to $\mathbb{B}$ gives

$$\frac{d}{d\mathbb{B}}\left(e^{\frac{\mathbb{B}}{2}} + e^{-\frac{\mathbb{B}}{2}}\right)^n = \frac{1}{2}n\left(e^{\frac{\mathbb{B}}{2}} + e^{-\frac{\mathbb{B}}{2}}\right)^n \left(\frac{e^{\frac{\mathbb{B}}{2}} - e^{-\frac{\mathbb{B}}{2}}}{e^{\frac{\mathbb{B}}{2}} + e^{-\frac{\mathbb{B}}{2}}}\right). \tag{A3.6}$$

Derivative of the right part of the eq. (A3.5) with respect to $\mathbb{B}$ gives

$$\frac{d}{d\mathbb{B}}\sum_{k=0}^{n} C_n^k e^{\left(\frac{n}{2}-k\right)\mathbb{B}} = \sum_{k=0}^{n} C_n^k \left(\frac{n}{2}-k\right) e^{\left(\frac{n}{2}-k\right)\mathbb{B}}. \tag{A3.7}$$

The right part of this equation is a numerator of the corresponding magnetization, e.g. $\frac{n}{2}-k$ corresponds to a spin projection, and $C_n^k$ is the coefficient of this spin projection's degeneracy, as detailed below.

## A4. Calculation of magnetization of arbitrary coupled spin-*I* systems

### Part 1: Proof of the Lemma 2 for *n* spin-½

We consider only one type of spins in the molecule: *n* is the number of equivalent spins per molecule, and *N* is the total number of molecules. Since the spins are magnetically equivalent, one has to consider the quantum mechanical rules of adding angular momenta. This results in the fact that total spin can take values starting from $J = nI$ down to zero for integer $J$ (**Fig. A1**) or $I$ for a half-integer $J$. At the high field, total magnetic moment can be computed via projections of the total spin using eq. (13):

$$m_z^{\{n\times I\},N} = N\gamma\hbar\cdot\left(\frac{\sum_{J_z=nI}^{-nI} J_z\cdot G_{n\times I}^{J_z}\cdot e^{\mathbb{B}J_z}}{\sum_{J_z=nI}^{-nI} G_{n\times I}^{J_z}\cdot e^{\mathbb{B}J_z}}\right) \xrightarrow{\text{HT }(\mathbb{B}\ll 1)} N\gamma\hbar\cdot\left(\mathbb{B}\frac{\sum_{J_z=nI}^{-nI} J_z^2\cdot G_{n\times I}^{J_z}}{\sum_{J_z=nI}^{-nI} G_{n\times I}^{J_z}}\right). \tag{A4.1}$$

Here, $G_{n\times I}^{J_z}$ are coefficients of a generalized Paskal triangle representing multiplicity (degeneracy) of the state with a projection $J_z$ of the maximal total spin $J = nI$.

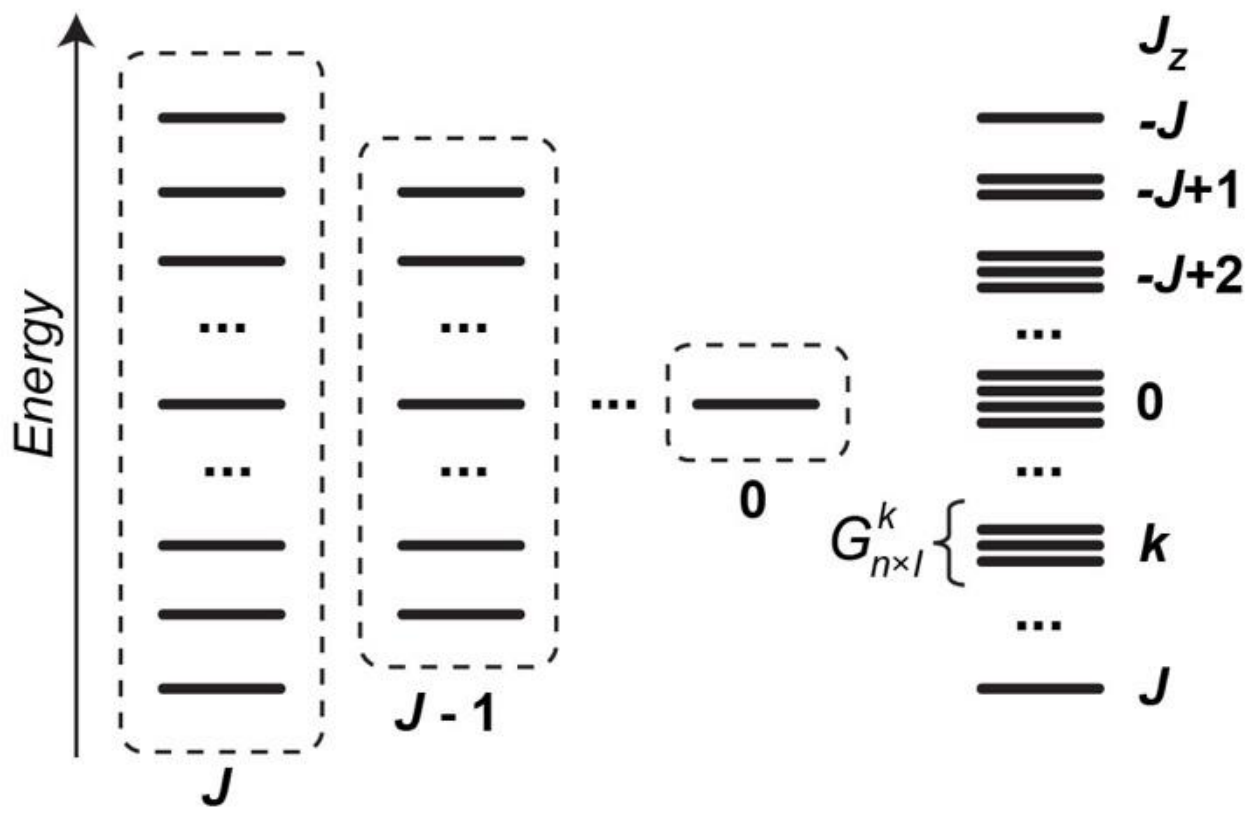

**Figure A1.** Illustration of the energy-level diagram for an *N*-spin system consisting of *n* spins-*I* per molecule. Left: manifolds corresponding to the total spin *J* = (*nI*), (*J* -1), … , 0 (the total spin-0 exists only when *nI* is an integer). Right: projections of the total spin and corresponding degeneracy of the energy levels, $G_{n\times I}^{J_\mathrm{z}}$.

Coefficients $G_{n\times I}^{J_\mathrm{z}}$ represent degeneracy of the energy level that corresponds to the total spin projection $J_\mathrm{z}$. For spin-½ systems, coefficients $G_{n\times\frac{1}{2}}^{J_\mathrm{z}}$ are given by the Pascal triangle [24,25], i.e., they are binomial coefficients $G_{n\times\frac{1}{2}}^{J_\mathrm{z}} = C_n^k$, where *n* is the number of spins, $k = n/2 - J_\mathrm{z}$ and conversely $J_\mathrm{z} = n/2 - k$. The total number of spin projections is therefore $(n+1)$.

Before finding the magnetization in the general case (eq A4.1), let us first consider this simpler case of $n$ spin-½ system with HT approximation. Thirst calculate part in brackets in eq (A4.1):

$$\left(\frac{\sum_{k=0}^{n} C_n^k \left(\frac{n}{2}-k\right) e^{\left(\frac{n}{2}-k\right)\mathbb{B}}}{\sum_{k=0}^{n} C_n^k e^{\left(\frac{n}{2}-k\right)\mathbb{B}}}\right) \approx \mathbb{B}\frac{\sum_{k=0}^{n} C_n^k \left(\frac{n}{2}-k\right)^2}{\sum_{k=0}^{n} C_n^k} = \mathbb{B}\frac{\frac{1}{4}n2^n}{2^n} = \frac{1}{4}n\mathbb{B} = \frac{1}{2}nP_\mathrm{z}^{\{1/2\},\mathrm{HT}}. \tag{A4.2}$$

Here, we used the results of eq. (A3.1) and (A3.4) and (3). Therefore eq. (A4.1) can be simplified as,

$$m_\mathrm{z}^{\{n\times 1/2\},N} = N\gamma\hbar\cdot\frac{1}{2}nP_\mathrm{z}^{\{1/2\},\mathrm{HT}} = n\cdot m_\mathrm{z}^{\{1/2\},N}. \tag{A4.3}$$

and the **Lemma 2** is proven for $n$ spin-½ system under HT approximation.

By using eqs. (A3.5-A3.7), without loss of generality for $n$ spin-½ system, part in brackets in eq. (A4.1) can be shown to be equal $\frac{1}{2}nP_\mathrm{z}^{1/2}$:

$$\frac{\sum_{k=0}^{n} C_n^k \left(\frac{n}{2}-k\right) e^{\left(\frac{n}{2}-k\right)\mathbb{B}}}{\sum_{k=0}^{n} C_n^k e^{\left(\frac{n}{2}-k\right)\mathbb{B}}} = \frac{\frac{1}{2}n\left(e^{\frac{\mathbb{B}}{2}}+e^{-\frac{\mathbb{B}}{2}}\right)^n \left(\frac{e^{\frac{\mathbb{B}}{2}}-e^{-\frac{\mathbb{B}}{2}}}{e^{\frac{\mathbb{B}}{2}}+e^{-\frac{\mathbb{B}}{2}}}\right)}{\left(e^{\frac{\mathbb{B}}{2}}+e^{-\frac{\mathbb{B}}{2}}\right)^n} = \frac{1}{2}n\cdot\tanh(\mathbb{B}/2) \ = \frac{1}{2}nP_\mathrm{z}^{\{\frac{1}{2}\}}. \tag{A4.4}$$

With this, **Lemma 2** is proven for $n$ spin-½ system at thermal equilibrium at any temperature.

Part 2: Proof of the Lemma 2 for *n* spin-*I*

Now, the same steps could be carried out for $n$ spin-*I* and find the magnetization $m_{\mathrm{z}}^{\{n\times I\},N}$ (eq A4.1). First, we know that the sum of all $G_{n\times I}^{J_{\mathrm{z}}}$ (denominator in eq A4.1) equals to the total number of states in the system:

$$\sum_{J_{\mathrm{z}}=nI}^{-nI} G_{n\times I}^{J_{\mathrm{z}}} = (2I+1)^n. \tag{A4.5}$$

Second, since these coefficients can be obtained by calculating the generalized Pascal triangle for spin-*I* systems [26], they represent relative intensities of lines in high-field NMR spectra of a spin coupled, for example, via *J*-coupling to a group of $n$ spin-*I*.

For example, for $n$ spin-½ these intensities are obtained from the following equation:

$$\left(e^{i\pi Jt} + e^{-i\pi Jt}\right)^n = \sum_{J_{\mathrm{z}}=-n/2}^{n/2} G_{n\times\frac{1}{2}}^{J_{\mathrm{z}}} e^{i\pi Jt J_{\mathrm{z}}}. \tag{A4.6}$$

Exponentials determine frequency offsets of lines in the spectrum, and the numbers before exponents will give us degeneration: for spin-½, coefficients $G$ coinside with binominal coefficients as discussed in previous appendices. Note that in this section $J$ represent spin-spin interaction and $J_{\mathrm{z}}$ is a projection of total spin but it is easer to see the difference as spin-spin interaction is always with $\pi$ on the left to it.

Now, for a general $AX_n$ spin system (where X are spin-*I* particles), the splittings are given by the evolution of the type (analogous to A4.6)

$$\left(\sum_{I_{\mathrm{z}}=-I}^{I} e^{i2\pi Jt I_{\mathrm{z}}}\right)^n = \sum_{J_{\mathrm{z}}=-In}^{In} G_{n\times I}^{J_{\mathrm{z}}} e^{i2\pi Jt J_{\mathrm{z}}} \tag{A4.7}$$

One can find $G_{n\times\frac{1}{2}}^{I\mathrm{z}}$ values and generalized Pascal triangles for any spin using multinomial coefficients but we do not need it here. Instead, we need to find only two sums: numerator and denominator in eq. (A4.1). We will start with a simpler case of HT approximation. To make notations shorter we will make a small substitution $i2\pi Jt \to a$. As before, first, we will calculate the denominator of eq. (A4.1) using eq. (A4.7) and putting $a \to 0$:

$$\left(\sum_{I_{\mathrm{z}}=-I}^{I} e^{aI_{\mathrm{z}}}\right)^n = \sum_{J_{\mathrm{z}}=-In}^{In} G_{n\times I}^{J_{\mathrm{z}}} e^{aI_{\mathrm{z}}} \xrightarrow{a\to 0} \left(\sum_{I_{\mathrm{z}}=-I}^{I} 1\right)^n = (2I+1)^n = \sum_{J_{\mathrm{z}}=-In}^{In} G_{n\times I}^{J_{\mathrm{z}}}. \tag{A4.8}$$

Now, we can find the numerator of eq. (A4.1). To do so, we will differentiate eq. (A4.8) twice and again put limits ($a \to 0$):

$$\begin{aligned}
&\frac{d^2}{da^2}\left(\sum_{I_{\mathrm{z}}=-I}^{I} e^{aI_{\mathrm{z}}}\right)^n = n\frac{d}{da}\left[\left(\sum_{I_{\mathrm{z}}=-I}^{I} I_{\mathrm{z}} e^{aI_{\mathrm{z}}}\right)\left(\sum_{I_{\mathrm{z}}=-I}^{I} e^{aI_{\mathrm{z}}}\right)^{n-1}\right] = \\
&= n\left(\sum_{I_{\mathrm{z}}=-I}^{I} I_{\mathrm{z}}^2 e^{aI_{\mathrm{z}}}\right)\left(\sum_{I_{\mathrm{z}}=-I}^{I} e^{aI_{\mathrm{z}}}\right)^{n-1} + n(n-1)\left(\sum_{I_{\mathrm{z}}=-I}^{I} I_{\mathrm{z}} e^{aI_{\mathrm{z}}}\right)^2\left(\sum_{I_{\mathrm{z}}=-I}^{I} e^{aI_{\mathrm{z}}}\right)^{n-2} \xrightarrow{a\to 0} \\
&n\left(\sum_{I_{\mathrm{z}}=-I}^{I} I_{\mathrm{z}}^2\right)\left(\sum_{I_{\mathrm{z}}=-I}^{I} 1\right)^{n-1} = n\frac{I(I+1)(2I+1)}{3}(2I+1)^{n-1} = n\frac{I(I+1)}{3}(2I+1)^n
\end{aligned} \tag{A4.9}$$

This equals to denominator of eq. (A4.1):

$$\frac{d^2}{da^2}\left(\sum_{I_{\mathrm{z}}=-I}^{I} e^{aI_{\mathrm{z}}}\right)^n = \frac{d^2}{da^2}\left(\sum_{J_{\mathrm{z}}=-In}^{In} J_{\mathrm{z}}^2 G_{n\times I}^{J_{\mathrm{z}}} e^{aJ_{\mathrm{z}}}\right) = \sum_{J_{\mathrm{z}}=-In}^{In} J_{\mathrm{z}}^2 G_{n\times I}^{J_{\mathrm{z}}} e^{aJ_{\mathrm{z}}} \xrightarrow{a\to 0} \sum_{J_Z=-In}^{In} J_{\mathrm{Z}}^2 G_{n\times I}^{J_{\mathrm{z}}} \tag{A4.10}$$

Hence, now we can combine the equations above and find the part in brackets for magnetization of *n* spins-*I* at HT approximation (eq. A4.1):

Here we used eq. A1.3 from **Appendix A1**. Hence

$$\left(\mathbb{B}\frac{\sum_{J_z=nI}^{-nI} J_z^2\cdot G_{n\times I}^{J_z}}{\sum_{J_z=nI}^{-nI} G_{n\times I}^{J_z}}\right) = \frac{n\frac{I(I+1)}{3}(2I+1)^n}{(2I+1)^n}\mathbb{B} = n\frac{I(I+1)}{3}\mathbb{B} = nIP_z^{\{I\},\mathrm{HT}}. \tag{A4.11}$$

This is precisely like eq. A4.2 but for general spin-*I*.

And finally, we can find eq. A4.1 and proof our Lemma 2 in general case without HT approximation. We will use the same ideas as before, but now everywhere, instead of parameter $a$, we will use $\mathbb{B}$ and easily proof the statement in the general case. The denominator of eq. A4.1 is:

$$\sum_{J_z=-nI}^{nI} G_{n\times I}^{J_z} \cdot e^{J_z\mathbb{B}} = \left(\sum_{I_z=-I}^{I} e^{I_z\mathbb{B}}\right)^n. \tag{A4.12}$$

And corresponding numerator is its first derivative:

$$\begin{aligned}\frac{d}{d\mathbb{B}}\sum_{J_z=-nI}^{nI} G_{n\times I}^{J_z} \cdot e^{J_z\mathbb{B}} &= \sum_{J_z=-nI}^{nI} G_{n\times I}^{J_z} \cdot J_z e^{J_z\mathbb{B}} = \\ n\sum_{I_z=-I}^{I} I_z e^{I_z\mathbb{B}}\left(\sum_{I_z=-I}^{I} e^{I_z\mathbb{B}}\right)^{n-1} &= nI\frac{\sum_{I_z=-I}^{I} I_z e^{I_z\mathbb{B}}}{I\sum_{I_z=-I}^{I} e^{I_z\mathbb{B}}}\left(\sum_{I_z=-I}^{I} e^{I_z\mathbb{B}}\right)^n = nIP_z^{\{I\}}\left(\sum_{I_z=-I}^{I} e^{I_z\mathbb{B}}\right)^n.\end{aligned} \tag{A4.13}$$

Here, we used the definition of $P_z^{\{I\}}$ (eq. 13). Now, combining eqs. (A5.8, A5.9 and A4.1) to get the following equation to prove **Lemma 1**.

$$m_z^{\{n\times I\},N} = N\gamma\hbar\cdot\left(\frac{\sum_{J_z=nI}^{-nI} J_z\cdot G_{n\times I}^{J_z}\cdot e^{\mathbb{B}J_z}}{\sum_{J_z=nI}^{-nI} G_{n\times I}^{J_z}\cdot e^{\mathbb{B}J_z}}\right) = N\cdot\gamma\hbar I\cdot n\cdot P_z^{\{I\}} = n\cdot m_z^{\{I\},N}. \tag{A4.14}$$

That is the quintessence of Lemma 2, and it is now proven in a general case here, demonstrating that magnetization is an additive quantity for equivalent spins of the same molecule.

## A5. Solution to the Problem 1

Magnetic moment corresponding to *N* spin-½ particles at HF is given by eqs. (8) and (13):

$$m_z^{\{1/2\},N} = N\cdot\left(\frac{\gamma\hbar}{2}\right)\cdot\left[\frac{e^{\frac{1}{2}\mathbb{B}}-e^{-\frac{1}{2}\mathbb{B}}}{e^{\frac{1}{2}\mathbb{B}}+e^{-\frac{1}{2}\mathbb{B}}}\right]. \tag{A5.1}$$

If now, instead, we have *N*/2 pairs of interacting spin-½ particles (the total number of spins is still *N*), we first need to change calculation basis accordingly. Two conventional basis sets are used to calculate the pairs of spins. The first is the Zeeman basis: $|1\rangle = |\alpha\alpha\rangle$, $|2\rangle = |\alpha\beta\rangle$, $|3\rangle = |\beta\alpha\rangle$, $|4\rangle = |\beta\beta\rangle$, and the second is the singlet-triplet (S-T) basis (**Figure 1B**): $|0,0\rangle = |S\rangle = (|\alpha\beta\rangle - |\beta\alpha\rangle)/\sqrt{2}$, $|1,1\rangle = |T_+\rangle = |\alpha\alpha\rangle$, $|1,0\rangle = |T_0\rangle = (|\alpha\beta\rangle + |\beta\alpha\rangle)/\sqrt{2}$, $|1,-1\rangle = |T_-\rangle = |\beta\beta\rangle$. Here, $|\alpha\rangle = |1/2\,,1/2\rangle$ and $|\beta\rangle = |1/2\,,-1/2\rangle$, are shorthand notations for spin-½ being parallel and antiparallel to the *z*-axis (here, the first number in the ket is the spin value, and the second number denotes its projection onto the *z*-axis). The S-T basis is a natural choice for two equivalent spins-½ because it separates spin states into two symmetrized manifolds with total spin of 1 and 0, respectively.

Although these two basis sets are different, the application of the eqs. (8-13) will still give the same result. This is because only projections $\mu_z(|\psi\rangle))$ of the total magnetic moment matter, and these projections are the same for Zeeman and S-T basis sets, i.e., {-1, 0, 0, 1}. Hence, one can calculate the magnetization of a pair of spin-½ particles as follows:

$$m_z^{\{2\times(1/2)\},N/2} = \frac{N}{2}\cdot(\gamma\hbar)\cdot\left[\frac{e^{\mathbb{B}}+0+0-e^{-\mathbb{B}}}{e^{\mathbb{B}}+1+1+e^{-\mathbb{B}}}\right] = N\cdot\left(\frac{\gamma\hbar}{2}\right)\cdot\left[\frac{e^{\frac{\mathbb{B}}{2}}-e^{-\frac{\mathbb{B}}{2}}}{e^{\frac{\mathbb{B}}{2}}+e^{-\frac{\mathbb{B}}{2}}}\right] = m_z^{\{1/2\},N}. \tag{A5.2}$$

Hence, given the same volume and sample geometry, magnetization of samples composed of equivalent spins (such as in $H_2$ or $H_2O$ molecules) and magnetization of samples composed of magnetically nonequivalent spins of the same concentration at thermal equilibrium and high field coincide. Our Theorem gives an extension to any number of equivalent spins. Note that this equivalence does not require HT approximation but requires thermodynamic equilibrium and HF conditions.

For spin-1 particles, magnetization is as follows

$$m_{\mathrm{z}}^{\{1\},N} = N \cdot (\gamma\hbar) \cdot \left[\frac{e^{\mathbb{B}} + 0 - e^{-\mathbb{B}}}{e^{\mathbb{B}} + 1 + e^{-\mathbb{B}}}\right]. \tag{A5.3}$$

One can see that the ratio of sample magnetizations as a function of spin quantum number for spin-½ and spin-1 particles is

$$\frac{m_{\mathrm{z}}^{\{1\},N}}{m_{\mathrm{z}}^{\{1/2\},N}} = 2 \cdot \left[1 + \frac{1}{e^{\mathbb{B}} + 1 + e^{-\mathbb{B}}}\right]. \tag{A5.4}$$

If $\mathbb{B} \gg 1$, $m_{\mathrm{z}}^{\{1\},N} / m_{\mathrm{z}}^{\{1/2\},N} \xrightarrow{\mathbb{B}\to\infty} 2$. Given HT approximation ($\mathbb{B} \ll 1$), however, $m_{\mathrm{z}}^{\{1\},N} / m_{\mathrm{z}}^{\{1/2\},N} \xrightarrow{\mathbb{B}\to 0} m_{\mathrm{z}}^{\{1\},N,\mathrm{HT}} / m_{\mathrm{z}}^{\{1/2\},N,\mathrm{HT}} = 8/3$.

### A6. Solution to the Problem 2

Fermi's golden rule determines transition probability per unit time between any two states $|1\rangle$ and $|2\rangle$ [27]:

$$W_{1\to 2} = \frac{2\pi}{\hbar} \left|\langle 2|\hat{V}_{\mathrm{RF}}|1\rangle\right|^2 (n_1 - n_2)\delta(E_2 - E_1 - \hbar\omega). \tag{A6.1}$$

Here, $n_1$ and $n_2$ are populations of the states with energies $E_1$ and $E_2$, respectively. The amplitudes of NMR spectral lines are proportional to the corresponding transition probabilities. In the following, we will neglect proportionality coefficients. An on-resonance RF-pulse induces the single-quantum transitions between spin states. The interaction of spins with the RF field is given by the following operator (in the rotation frame of reference):

$$\hat{V}_{\mathrm{RF}} = \gamma\hbar B_1 \hat{I}_{\mathrm{x}}. \tag{A6.2}$$

For the calculation of intensities, we will use lowering and raising spin operators $\hat{I}_-$ and $\hat{I}_+$ [28]:

$$\hat{I}_{\mathrm{x}} = \frac{\hat{I}_+ + \hat{I}_-}{2}, \hat{I}_{\mathrm{y}} = \frac{\hat{I}_+ - \hat{I}_-}{2i}. \tag{A6.3}$$

One can find the effect of these operators on spin states with total spin $I$ and projection $I_{\mathrm{z}}$:

$$\begin{aligned} \hat{I}_+|I, I_{\mathrm{z}}\rangle &= \sqrt{(I - I_{\mathrm{z}})(I + I_{\mathrm{z}} + 1)}|I, I_{\mathrm{z}} + 1\rangle, \\ \hat{I}_-|I, I_{\mathrm{z}}\rangle &= \sqrt{(I + I_{\mathrm{z}})(I - I_{\mathrm{z}} + 1)}|I, I_{\mathrm{z}} - 1\rangle. \end{aligned} \tag{A6.4}$$

Note that $\hat{I}_+|I, I\rangle = 0$ and $\hat{I}_-|I, -I\rangle = 0$. Now we are ready to use the eq. (A6.1) and apply it to the specific cases.

**<u>One spin-½ (A)</u>**. First, one can see that "up" and "down" transitions for one spin-½ (A-system) have the same intensity governed by the following matrix elements:

$$\langle\alpha|\hat{I}_{\mathrm{x}}|\beta\rangle = \langle\beta|\hat{I}_{\mathrm{x}}|\alpha\rangle = \left\langle\beta\left|\frac{\hat{I}_-}{2}\right|\alpha\right\rangle = 1/2. \tag{A6.5}$$

Therefore, the NMR signal of the A-system consisting of *N* particles is

$$\mathrm{Signal}(N,\mathrm{A}) \sim N \cdot W_{\alpha\to\beta} \sim \left(\frac{\hbar\gamma B_1}{2}\right)^2 \left(n_\alpha - n_\beta\right) = N\left(\frac{\hbar\gamma B_1}{2}\right)^2 \left[\frac{e^{\frac{1}{2}\mathbb{B}} - e^{-\frac{1}{2}\mathbb{B}}}{e^{\frac{1}{2}\mathbb{B}} + e^{-\frac{\mathbb{B}}{2}}}\right] \xrightarrow{\mathbb{B}\to 0} N(\hbar\gamma B_1)^2 \cdot \frac{1}{8}\mathbb{B}. \quad \text{(A6.5)}$$

**<u>Two spins-½ (A₂)</u>**. In the system of two equivalent spins ($A_2$-system), there are four single quantum transitions, but only two of them are allowed:

$$\langle T_- | \hat{I}_\mathrm{x} | T_0 \rangle = \langle T_0 | \hat{I}_\mathrm{x} | T_+ \rangle = \left\langle T_0 \left| \frac{\hat{I}_-}{2} \right| T_+ \right\rangle = 1/\sqrt{2}, \quad \text{(A6.6)}$$
$$\langle T_{+1} | \hat{I}_\mathrm{x} | S \rangle = \langle T_{-1} | \hat{I}_\mathrm{x} | S \rangle = 0.$$

Here, $\hat{I}_\mathrm{x} = \hat{I}_{1\mathrm{x}} + \hat{I}_{2\mathrm{x}}$ is the operator of the *x*-projection of the total spin. The NMR signal from *N*/2 pairs of equivalent spins is a sum of two signals corresponding to the two allowed transitions:

$$\mathrm{Signal}(N/2,\mathrm{A}_2) \sim \frac{N}{2}\left(W_{T_0\to T_-} + W_{T_+\to T_0}\right) \sim \frac{1}{2}\left(\frac{\gamma\hbar B_1}{\sqrt{2}}\right)^2 \left[\left(n_{T_0} - n_{T_-}\right) + \left(n_{T_+} - n_{T_0}\right)\right] = \quad \text{(A6.7)}$$
$$= \frac{N}{4}(\hbar\gamma B_1)^2 \left[\frac{e^{\mathbb{B}} - e^{-\mathbb{B}}}{e^{\mathbb{B}} + 1 + 1 + e^{-\mathbb{B}}}\right] = N\left(\frac{\hbar\gamma B_1}{2}\right)^2 \left[\frac{e^{\mathbb{B}/2} - e^{-\mathbb{B}/2}}{e^{\mathbb{B}/2} + e^{-\mathbb{B}/2}}\right] = \mathrm{Signal}(N,\mathrm{A}).$$

Once again, one can see that the signal from *N*/2 of $A_2$-systems is the same as from *N* of A-systems.

**<u>Two spins-½ (AX)</u>**. In the case of two chemically nonequivalent spins (AX-system), there are four single-quantum transitions, all of which have the same intensity:

$$\langle \alpha\beta | \hat{I}_\mathrm{x} | \alpha\alpha \rangle = \langle \beta\alpha | \hat{I}_\mathrm{x} | \alpha\alpha \rangle = \langle \beta\beta | \hat{I}_\mathrm{x} | \alpha\beta \rangle = \langle \beta\beta | \hat{I}_\mathrm{x} | \beta\alpha \rangle = \frac{1}{2}, \quad \text{(A6.8)}$$

where again, $\hat{I}_\mathrm{x} = \hat{I}_{1\mathrm{x}} + \hat{I}_{2\mathrm{x}}$. Thus, the total NMR signal from *N*/2 pairs of nonequivalent spins (AX-systems) is a sum of four signals from the four allowed transitions:

$$\mathrm{Signal}(N/2,\mathrm{AX}) \sim \frac{N}{2}\left(W_{\alpha\alpha\to\alpha\beta} + W_{\alpha\alpha\to\beta\alpha} + W_{\alpha\beta\to\beta\beta} + W_{\beta\alpha\to\beta\beta}\right) \sim \quad \text{(A6.9)}$$
$$\sim \frac{1}{2}\left(\frac{\gamma\hbar B_1}{2}\right)^2 \left[\left(n_{\alpha\alpha} - n_{\alpha\beta}\right) + \left(n_{\alpha\alpha} - n_{\beta\alpha}\right) + \left(n_{\alpha\beta} - n_{\beta\beta}\right) + \left(n_{\beta\alpha} - n_{\beta\beta}\right)\right] =$$
$$= \frac{N}{4}(\hbar\gamma B_1)^2 \left[\frac{e^{\mathbb{B}} - e^{-\mathbb{B}}}{e^{\mathbb{B}} + 1 + 1 + e^{-\mathbb{B}}}\right] = N\left(\frac{\hbar\gamma B_1}{2}\right)^2 \left[\frac{e^{\mathbb{B}/2} - e^{-\mathbb{B}/2}}{e^{\mathbb{B}/2} + e^{-\mathbb{B}/2}}\right] = \mathrm{Signal}(N,\mathrm{A}).$$

One can see that the signal from *N*/2 of AX-systems, N/2 of $A_2$-system, and *N* of A-systems is the same. This is another demonstration of the validity of the Theorem, and confirmation of the fact that groupling of spins does not change their observable signal at HF.

**<u>One spin-1 (A)</u>.** In the system of one spin-1 particles (A-system), there are only two single-quantum transitions, and they have the same intensity:

$$\langle 1,-1 | \hat{I}_\mathrm{x} | 1,0 \rangle = \langle 1,0 | \hat{I}_\mathrm{x} | 1,1 \rangle = \left\langle 1,0 \left| \frac{\hat{I}_-}{2} \right| 1,1 \right\rangle = \frac{1}{\sqrt{2}}. \quad \text{(A6.10)}$$

This case is very similar to the case of two equivalent spins-½ (eq. A6.7) except for the lack of the singlet state. Thus, the total NMR signal from *N* spins-1 is a sum of two signals from the two allowed transitions:

$$\mathrm{Signal}(N,A) \sim N \cdot (W_{+1\to 0} + W_{0\to -1}) \sim \left(\frac{\gamma\hbar B_1}{\sqrt{2}}\right)^2 \left[\,(n_{+1} - n_0) + (n_0 - n_{-1})\right] = \quad \text{(A6.11)}$$
$$= \frac{N}{2}(\gamma\hbar B_1)^2 \left[\frac{e^{\mathbb{B}} - e^{-\mathbb{B}}}{e^{\mathbb{B}} + 1 + e^{-\mathbb{B}}}\right] \xrightarrow{\mathbb{B}\to 0} N(\hbar\gamma B_1)^2 \cdot \frac{1}{3}\mathbb{B}.$$

Thus, the receptivity of NMR detection at HF concerning spins-1 is 8/3 higher than that of spins-½, considering the same gyromagnetic ratio and spin number density. This result also matches the ratio of magnetizations obtained before (eq. A5.4).

## A7. Solution to the Problem 3

We will first consider HT approximation case, and then a general case.

*1. HT approximation.*

Magnetization of *N* spins-*I* at thermal equilibrium is

$$m_z^{\{I\},N} = N \cdot (\gamma\hbar I)\frac{I+1}{3}\mathbb{B}. \quad \text{(A7.1)}$$

From the **Theorem 1** we can find the magnetization of a pair of spins:

$$m_z^{\left\{\frac{1}{2},\frac{1}{2}\right\},N} = 2m_z^{\left\{\frac{1}{2}\right\},N} = \frac{1}{2}N\gamma\hbar\mathbb{B}. \quad \text{(A7.2)}$$

For two-spin-½ systems at HT, all states are almost equally populated; hence, ¾N quasi-particles have spin-1 (triplet spin pairs), and ¼*N* have spin-0 (singlet spin pairs). Now lets calculate magnetization as a superposition of these two ansembles of particles:

$$m_z^{\{1\},\frac{3}{4}N,\mathrm{HT}} + m_z^{\{0\},\frac{1}{4}N,\mathrm{HT}} = \frac{3}{4}m_z^{\{1\},N,\mathrm{HT}} = \frac{3}{4}N\frac{2\gamma\hbar}{3}\mathbb{B} = \frac{1}{2}N\gamma\hbar\mathbb{B}. \quad \text{(A7.3)}$$

Hence, magnetization in this case can be calculated as a superposition of quasi-particles. Interestingly, the loss of total magnetization due to the population of the unobservable singlet states is compensated by the "stronger-magnetized" triplet manifolds.

*2. General case.* Let us now consider the general case of arbitrary numbers of equivalent spins-*I*. These spins can be grouped in a set of states associated with the total spin $J$ such that each state from the set can be obtained using the rising or lowering operators of total spin (**Figure A1**). We will call each such a set of states a quasi-particle with spin-$J$ as it can be described by its own orthogonal basis independent from other sets and it has all spin projections from $J$ to $-J$. By definition, magnetization of the system is as follows:

$$m_z^{\{I_1 I_2 \dots\},N} = \frac{1}{\mathbb{Z}}N\gamma\hbar\sum_{\{I_{1\mathrm{z}}I_{2\mathrm{z}}\dots\}}(I_{1\mathrm{z}} + I_{2\mathrm{z}} \dots)e^{(I_{1\mathrm{z}}+I_{2\mathrm{z}}\dots)\mathbb{B}}. \quad \text{(A7.4)}$$

Here, the summation goes over all combinations of spin projections of individual spins. Using, however, the total spin basis, it can be written as

$$m_z^{\{I_1 I_2 \dots\},N} = \frac{1}{\mathbb{Z}}N\gamma\hbar\sum_J\sum_{J_\mathrm{z}=J}^{-J}J_\mathrm{z}e^{J_\mathrm{z}\mathbb{B}}. \quad \text{(A7.5)}$$

The last two equations are identical and differ only by how spin states are grouped. The equality (A.7.5) already gives a hint to the problem's solution. The statistical sum for each quasiparticle with spin-$J$ is $\mathbb{Z}_J = \sum_{J_z=J}^{-J}e^{J_z\mathbb{B}}$. Let us now introduce it in the eq. (A7.5) as follows:

$$m_z^{\{I_1 I_2 \dots\},N} = \sum_J\frac{\mathbb{Z}_J}{\mathbb{Z}}N\frac{\sum_{J_\mathrm{z}=J}^{-J}\gamma\hbar J_\mathrm{z}e^{J_\mathrm{z}\mathbb{B}}}{\mathbb{Z}_J} = \sum_J\frac{\mathbb{Z}_J}{\mathbb{Z}}m_\mathrm{z}^{\{J_\mathrm{z}\},N} = \sum_J\rho_J m_\mathrm{z}^{\{J_\mathrm{z}\},N} = \sum_J m_\mathrm{z}^{\{J_\mathrm{z}\},N\rho_J}. \quad \text{(A7.6)}$$

Here, $\rho_J = \mathbb{Z}_J/\mathbb{Z}$ is a statistical fraction of realizing a particular total spin-*J* out of all possible total spins. Hence, indeed, one can calculate the magnetization of equivalent spins as a superposition of magnetizations comprising pseudo particles.

### A8. Calculation of polarization of arbitrary coupled spin-*I* systems. Part 3: Proof of the equivalent formulation of the theorem (eq. 32)

In the main text, we derived $P_{rz}^{\{I_1 I_2 \ldots I_n\},\mathrm{HT}}$ (eq. 31). Now lets find it without HT approximation and demonstrate that $P_{rz}^{\{I_1 I_2 \ldots I_n\}} = P_{\mathrm{z}}^{\{I_r\}}$ at thermal equilibrium. Polarization of spin-$I_r$ as in a group of *n*-spins is

$$P_{rz}^{\{I_1 I_2 \ldots I_n\}} = \frac{1}{I_r}\mathrm{Tr}\left(\hat{I}_{rz}^{\{I_1 I_2 \ldots I_n\}}\hat{\rho}^{\{I_1 I_2 \ldots I_n\}}\right) = \frac{1}{I_r}\frac{\mathrm{Tr}\left(\hat{I}_{rz}^{\{I_1 I_2 \ldots I_n\}}\prod_m e^{\mathbb{B}^m \hat{I}_{mz}^{\{I_1 I_2 \ldots I_n\}}}\right)}{\mathrm{Tr}\left(\prod_m e^{\mathbb{B}^m \hat{I}_{mz}^{\{I_1 I_2 \ldots I_n\}}}\right)} \tag{A8.1}$$

When working with density matrices following equalities are very useful and will be used below often:

$$\mathrm{Tr}(A \otimes B) = \mathrm{Tr}(A)\mathrm{Tr}(B), \tag{A8.2}$$

for any square matrices $A$ and $B$. Additionally we needed the mixed-product property of the direct product operator:

$$(A \otimes B) \cdot (C \otimes D) = A \cdot C \otimes B \cdot D. \tag{A8.3}$$

where A, B, C, and D are matrices of the appropriate size for scalar (dot) product.

Lets simplify denominator in eq. A8.1:

$$\begin{aligned}\mathrm{Tr}\left(\prod_m e^{\mathbb{B}^m \hat{I}_{mz}^{\{I_1 I_2 \ldots I_n\}}}\right) &= \mathrm{Tr}\left(\prod_m e^{\mathbb{B}^m (\hat{1}_1 \otimes \ldots \hat{1}_{\mathrm{r}-1} \otimes \hat{I}_{mz} \otimes \hat{1}_{\mathrm{r}+1} \ldots \otimes \hat{1}_n)}\right) = \\ &= \mathrm{Tr}\left(\bigotimes_{\mathrm{m}} e^{\mathbb{B}^m \hat{I}_{mz}}\right) = \prod_{m=1}^{N} \mathrm{Tr}\left(e^{\mathbb{B}^m \hat{I}_{mz}}\right).\end{aligned} \tag{A8.4}$$

Nominator equals to

$$\begin{aligned}&\mathrm{Tr}\left(\hat{I}_{rz}^{\{I_1 I_2 \ldots I_n\}}\prod_m e^{\mathbb{B}^m \hat{I}_{mz}^{\{I_1 I_2 \ldots I_n\}}}\right) = \frac{d}{d(\mathbb{B}^r)}\mathrm{Tr}\left(\prod_m e^{\mathbb{B}^m \hat{I}_{mz}^{\{I_1 I_2 \ldots I_n\}}}\right) = \frac{d}{d(\mathbb{B}^r)}\prod_{m=1}^{N}\mathrm{Tr}\left(e^{\mathbb{B}^m \hat{I}_{mz}}\right) = \\ &= \frac{\prod_{m=1}^{N}\mathrm{Tr}\left(e^{\mathbb{B}^m \hat{I}_{mz}}\right)}{\mathrm{Tr}\left(e^{\mathbb{B}^r \hat{I}_{rz}}\right)}\frac{d\mathrm{Tr}\left(e^{\mathbb{B}^r \hat{I}_{rz}}\right)}{d(\mathbb{B}^r)} = \frac{\prod_{m=1}^{N}\mathrm{Tr}\left(e^{\mathbb{B}^m \hat{I}_{mz}}\right)}{\mathrm{Tr}\left(e^{\mathbb{B}^r \hat{I}_{rz}}\right)}\mathrm{Tr}\left(\hat{I}_{rz} e^{\mathbb{B}^r \hat{I}_{rz}}\right).\end{aligned} \tag{A8.5}$$

From eq. (4.1-3) it follows that

$$P_{rz}^{\{I_1 I_2 \ldots I_n\}} = \frac{1}{I_r}\frac{\mathrm{Tr}\left(\hat{I}_{rz}^{\{I_1 I_2 \ldots I_n\}}\prod_m e^{\mathbb{B}^m \hat{I}_{mz}^{\{I_1 I_2 \ldots I_n\}}}\right)}{\mathrm{Tr}\left(\prod_m e^{\mathbb{B}^m \hat{I}_{mz}^{\{I_1 I_2 \ldots I_n\}}}\right)} = \frac{\mathrm{Tr}\left(\hat{I}_{rz} e^{\mathbb{B}^r \hat{I}_{rz}}\right)}{\mathrm{Tr}\left(e^{\mathbb{B}^r \hat{I}_{rz}}\right)} = P_{\mathrm{z}}^{\{I_r\}}. \tag{A8.6}$$

Hence, the statement was proved: the polarization of a spin is system-size independent because, on the right-hand side, there is no information regarding the size of the system at HF-approximation.

## A9. Quadrupolar polarization

To find “quadrupolar polarization” (eq. 41), $\hat{P}_{\mathrm{Q}}^{\{I\}} = \frac{3(\hat{I}_{\mathrm{z}})^2 - I(I+1)\hat{1}}{I^2}$, we need to know the following quantity: $\mathrm{Tr}\left(\left(\hat{I}_{\mathrm{z}}\right)^2 e^{\mathbb{B}\hat{I}_{\mathrm{z}}}\right)$ and $\mathrm{Tr}\left(e^{\mathbb{B}\hat{I}_{\mathrm{z}}}\right)$. We will use the same approach as before in appendix 1. We already deried $\mathrm{Tr}\left(e^{\mathbb{B}\hat{I}_{\mathrm{z}}}\right)$ (eq A1.1) and $\frac{d}{d\mathbb{B}}\mathrm{Tr}\left(e^{\mathbb{B}\hat{I}_{\mathrm{z}}}\right) = \mathrm{Tr}\left(\hat{I}_{\mathrm{Z}} e^{\mathbb{B}\hat{I}_{\mathrm{z}}}\right)$ (eq A1.2) and now we will find $\mathrm{Tr}\left(\left(\hat{I}_{\mathrm{z}}\right)^2 e^{\mathbb{B}\hat{I}_{\mathrm{z}}}\right)$:

$$\frac{d^2}{d\mathbb{B}^2}\mathrm{Tr}\left(e^{\mathbb{B}\hat{I}_{\mathrm{z}}}\right) = \mathrm{Tr}\left(\left(\hat{I}_{\mathrm{z}}\right)^2 e^{\mathbb{B}\hat{I}_{\mathrm{z}}}\right) = \tag{A9.1}$$
$$= \frac{e^{\left(I+\frac{1}{2}\right)\mathbb{B}} - e^{-\left(I+\frac{1}{2}\right)\mathbb{B}}}{e^{\frac{\mathbb{B}}{2}} - e^{-\frac{\mathbb{B}}{2}}}\left(\left(I+\frac{1}{2}\right)^2 + \frac{1}{4} + \frac{1}{2\sinh^2\left(\frac{\mathbb{B}}{2}\right)}\right) - \left(I+\frac{1}{2}\right)\frac{e^{\left(I+\frac{1}{2}\right)\mathbb{B}} + e^{-\left(I+\frac{1}{2}\right)\mathbb{B}}}{e^{\frac{\mathbb{B}}{2}} - e^{-\frac{\mathbb{B}}{2}}}\coth\left(\frac{\mathbb{B}}{2}\right).$$

Therefore, combining eq A1.1, A1.2, and eq. 41 quadrupolar polarization can be found

$$P_{\mathrm{Q}}^{\{I\}} = \frac{3\mathrm{Tr}\left((\hat{I}_{\mathrm{z}})^2 e^{\mathbb{B}\hat{I}_{\mathrm{z}}}\right) - I(I+1)}{I^2\mathrm{Tr}\left(e^{\mathbb{B}\hat{I}_{\mathrm{z}}}\right)} = \frac{2(I+1)}{I} + \frac{3}{2I^2} + \frac{3}{2I^2\sinh^2\left(\frac{\mathbb{B}}{2}\right)} - \frac{3}{I^2}\left(I+\frac{1}{2}\right)\coth\left(\left(I+\frac{1}{2}\right)\mathbb{B}\right)\coth\left(\frac{\mathbb{B}}{2}\right). \tag{A9.2}$$

For example, for spin-1 particles, the quadrupolar polarization is

$$P_{\mathrm{Q}}^{\{1\}} = \frac{11}{2} + \frac{3}{2\sinh^2\left(\frac{\mathbb{B}}{2}\right)} - \frac{9}{2}\coth\left(\frac{3}{2}\mathbb{B}\right)\coth\left(\frac{\mathbb{B}}{2}\right). \tag{A9.3}$$

## A10. Constraints on the elements of the density matrix of two spin-½

Let us look at the diagonal elements in density matrix, $\hat{\rho}^{\{½½\}}$, in Zeeman basis when only net magnetization of each spin and a two-spin order (eq. 37) are present in the system. All diagonal elements (populations) must be real numbers between 0 and 1:

$$\begin{aligned}
1 &\geq \rho_{\alpha\alpha} = \frac{1}{4} + \frac{1}{4}P_{1\mathrm{z}}^{\{½\}} + \frac{1}{4}P_{2\mathrm{z}}^{\{½\}} + \frac{1}{4}P_{\mathrm{zz}}^{\{½½\}} \geq 0, \\
1 &\geq \rho_{\alpha\beta} = \frac{1}{4} + \frac{1}{4}P_{1\mathrm{z}}^{\{½\}} - \frac{1}{4}P_{2\mathrm{z}}^{\{½\}} - \frac{1}{4}P_{\mathrm{zz}}^{\{½½\}} \geq 0, \\
1 &\geq \rho_{\beta\alpha} = \frac{1}{4} - \frac{1}{4}P_{1\mathrm{z}}^{\{½\}} + \frac{1}{4}P_{2\mathrm{z}}^{\{½\}} - \frac{1}{4}P_{\mathrm{zz}}^{\{½½\}} \geq 0, \\
1 &\geq \rho_{\beta\beta} = \frac{1}{4} - \frac{1}{4}P_{1\mathrm{z}}^{\{½\}} - \frac{1}{4}P_{2\mathrm{z}}^{\{½\}} + \frac{1}{4}P_{\mathrm{zz}}^{\{½½\}} \geq 0.
\end{aligned} \tag{A10.1}$$

These 8 inequalities can be written in the matrix form as

$$3 \geq \begin{pmatrix} +1 & +1 & +1 \\ +1 & -1 & -1 \\ -1 & +1 & -1 \\ -1 & -1 & +1 \end{pmatrix}\begin{pmatrix} P_{1\mathrm{z}}^{\{½\}} \\ P_{2\mathrm{z}}^{\{½\}} \\ P_{\mathrm{zz}}^{\{½½\}} \end{pmatrix} \geq -1. \tag{A10.2}$$

All conditions “3 ≥” are fulfilled automatically because all polarization values are between -1 and 1. Sometimes it is useful to quantify polarization as an average net polarization, $P_{\mathrm{net}} = \frac{P_{1\mathrm{z}}^{\{½\}} + P_{2\mathrm{z}}^{\{½\}}}{2}$, antiphase polarization, $P_{\mathrm{anti}} = \frac{P_{1\mathrm{z}}^{\{½\}} - P_{2\mathrm{z}}^{\{½\}}}{2}$, and zz-spin order, $P_{\mathrm{zz}} = P_{\mathrm{zz}}^{\{½½\}}$ , of the two-spin system. Then, the density matrix and restrictions can be written as

$$\begin{gathered}
\hat{\rho}^{\{½½\}} = \frac{\hat{1}}{4} + \frac{1}{2}P_{\mathrm{net}}\left(\hat{I}_{1\mathrm{z}}^{\{½½\}} + \hat{I}_{2\mathrm{z}}^{\{½½\}}\right) + \frac{1}{2}P_{\mathrm{anti}}\left(\hat{I}_{1\mathrm{z}}^{\{½½\}} - \hat{I}_{2\mathrm{z}}^{\{½½\}}\right) + P_{\mathrm{zz}}\hat{I}_{1\mathrm{z}}^{\{½½\}}\hat{I}_{2\mathrm{z}}^{\{½½\}}, \\
\begin{pmatrix} 2 & 0 & +1 \\ 0 & 2 & -1 \\ 0 & -2 & -1 \\ -2 & 0 & +1 \end{pmatrix}\begin{pmatrix} P_{\mathrm{net}} \\ P_{\mathrm{anti}} \\ P_{\mathrm{zz}} \end{pmatrix} \geq -1.
\end{gathered} \tag{A10.3}$$

From eq. (A10.3) one can get the following 8 conditions (**Figure 5A**):

$$1 \geq 0.5P_{zz} + 0.5 \geq P_{net} \geq -0.5P_{zz} - 0.5 \geq -1,$$
$$1 \geq -0.5P_{zz} + 0.5 \geq P_{anti} \geq 0.5P_{zz} - 0.5 \geq -1, \quad \text{(A10.4)}$$
$$1 \geq 1 - 2|P_{anti}| \geq P_{zz} \geq -1 + 2|P_{net}| \geq -1.$$

One can now see that there are restrictions on $P_{anti}$ and $P_{net}$ to be less than 1, the sum of their absolute values should be also smaller than one: $1 \geq |P_{net}| + |P_{anti}|$.

If $P_{anti}$ = 0 then eq. A11.4 (**Figure 4B**) simplifies to

$$1 \geq 0.5P_{zz} + 0.5 \geq P_{net} \geq -0.5P_{zz} - 0.5 \geq -1, \quad \text{(A10.5)}$$
$$1 \geq P_{zz} \geq -1 + 2|P_{net}| \geq -1.$$